\documentclass[12pt,preprint]{aastex}
\usepackage{amsmath}
\renewcommand {\ga}    {\mbox{\rlap{\hbox{\lower5pt\hbox{$\sim$}}}\hbox{$>$}}}
\renewcommand {\la}    {\mbox{\rlap{\hbox{\lower5pt\hbox{$\sim$}}}\hbox{$<$}}}

\begin{document}



\def\kms {\hbox{km{\hskip0.1em}s$^{-1}$}} 
 \def\ee #1 {\times 10^{#1}}          
 \def\ut #1 #2 { \, \mathrm{#1}^{#2}} 
 \def\u #1 { \, \mathrm{#1}}          
\def\lsol{\hbox{$\hbox{L}_\odot$}}
\def\Blos{B$_{\rm los}$}
\def\etal   {{\it et al. }}                     
\def\psec           {$.\negthinspace^{s}$}
\def\pasec          {$.\negthinspace^{\prime\prime}$}
\def\pdeg           {$.\kern-.25em ^{^\circ}$}
\def\degree{\ifmmode{^\circ} \else{$^\circ$}\fi}
\def\ee #1 {\times 10^{#1}}          
\def\ut #1 #2 { \, \textrm{#1}^{#2}} 
\def\u #1 { \, \textrm{#1}}          
\def\nH {n_\mathrm{H}}
\def\ddeg   {\hbox{$.\!\!^\circ$}}              
\def\deg    {$^{\circ}$}                        
\def\le     {$\leq$}                            
\def\sec    {$^{\rm s}$}                        
\def\msol   {\hbox{M$_\odot$}}                  
\def\i      {\hbox{\it I}}                      
\def\v      {\hbox{\it V}}                      
\def\dasec  {\hbox{$.\!\!^{\prime\prime}$}}     
\def\asec   {$^{\prime\prime}$}                 
\def\dasec  {\hbox{$.\!\!^{\prime\prime}$}}     
\def\dsec   {\hbox{$.\!\!^{\rm s}$}}            
\def\min    {$^{\rm m}$}                        
\def\hour   {$^{\rm h}$}                        
\def\amin   {$^{\prime}$}                       
\def\lsol{\, \hbox{$\hbox{L}_\odot$}}
\def\sec    {$^{\rm s}$}                        
\def\etal   {{\it et al. }}                     
\def\xbar   {\hbox{$\overline{\rm x}$}}         
\newcommand\percc{\textrm{cm}^{-3}}
\def\ee #1 {\times 10^{#1}}          
\def\ut #1 #2 { \, \textrm{#1}^{#2}} 
\def\un #1 { \, \textrm{#1}}          

\def\kms {\,\textrm{km\,s}^{-1}}
\def\persec {\, \hbox{s}^{-1}}
\def\percc {\,\mathrm{cm}^{-3}}
\def\la{\lower.4ex\hbox{$\;\buildrel <\over{\scriptstyle\sim}\;$}}
\def\ga{\lower.4ex\hbox{$\;\buildrel >\over{\scriptstyle\sim}\;$}}
\def\refitem{\par\noindent\hangindent\parindent}
\def\micron{\,$\mu$m}
\newcommand\topic[2]{\section*{#1 {\rm\small (#2)}}}

\newcommand{\agestar}{t_{\star}}
\newcommand{\mstar}{M_{\star}}
\newcommand{\rstar}{R_{\star}}
\newcommand{\tstar}{T_{\star}}
\newcommand{\lstar}{L_{\star}}
\newcommand{\mdote}{\dot{M}_{\rm env}}
\newcommand{\rmaxe}{R_{\rm env}^{\rm max}}
\newcommand{\rmine}{R_{\rm env}^{\rm min}}
\newcommand{\mdisk}{M_{\rm disk}}
\newcommand{\mdotdisk}{\dot{M}_{\rm disk}}
\newcommand{\rmaxd}{R_{\rm disk}^{\rm max}}
\newcommand{\rmind}{R_{\rm disk}^{\rm min}}
\newcommand{\rsub}{R_{\rm sub}}
\newcommand{\rhoconst}{\rho_{\rm cavity}}
\newcommand{\thetacav}{\theta_{\rm cavity}}

\newcommand{\um}{$\mu$m}
\newcommand{\cm}{cm$^{-1}$}
\newcommand{\vlsr}{v$_{LSR}$}
\newcommand{\palpha}{P\,$\alpha$}

\shorttitle{}
\shortauthors{}

\title{An Inverse Compton Scattering Origin of X-ray \\
Flares from Sgr A*}
\author{F. Yusef-Zadeh$^1$, M. Wardle$^2$, K. Dodds-Eden$^3$, C. O. Heinke$^4$, S. Gillessen$^3$, R. Genzel$^3$,  H. 
Bushouse$^5$, N. Grosso$^6$ \& D. Porquet$^6$}
\affil{$^1$Department of Physics and Astronomy, Northwestern University, Evanston, IL 60208}
\affil{$^2$Department of Physics and Astronomy, Macquarie University, Sydney NSW 2109, Australia}
\affil{$^3$Max Planck Institut f\"ur Extraterrestrische Physik, Postfach 1312, D-85741 Garching, Germany}
\affil{$^4$Dept. of Physics, University of Alberta, 4-183 CCIS, Edmonton AB T6G 2E1, Canada}
\affil{$^5$Space Telescope Science Institute, 3700 San Martin Drive, Baltimore, MD  21218}
\affil{$^6$Observatoire astronomique de Strasbourg, Universit\'e
de Strasbourg, CNRS, INSU, 11 rue de l'Universit\'e, 67000 Strasbourg, France}

\begin{abstract} 

The X-ray and near-IR emission from Sgr A* is dominated by flaring, while a quiescent component dominates 
the emission at radio and sub-mm wavelengths. The spectral energy distribution of the quiescent emission 
from Sgr A* peaks at sub-mm wavelengths and is modeled as synchrotron radiation from a thermal population of 
electrons in the accretion flow, with electron temperatures ranging up to $\sim 5-20$\,MeV.  Here we 
investigate the mechanism by which X-ray flare emission is produced through the interaction of the quiescent
and flaring components of Sgr A*. The X-ray flare emission has been interpreted as inverse Compton, 
self-synchrotron-Compton, or synchrotron emission.  We present results of simultaneous X-ray and near-IR 
observations and show evidence that X-ray peak flare emission lags behind near-IR flare emission with a 
time delay ranging from a few to tens of minutes.  Our Inverse Compton scattering modeling places constraints 
on the electron density and temperature distributions of the accretion flow and on the locations where flares
are produced. In the context of this model, the strong X-ray counterparts to near-IR flares arising from
the inner disk should show no significant time delay, whereas near-IR flares in the outer disk should show a 
broadened and delayed X-ray flare. 

\end{abstract}


\section{Introduction}

Observations of stellar orbits in the proximity of the enigmatic radio source Sgr~A*, located at the dynamical 
center of our galaxy, have shown compelling evidence that it is associated with a 4 $\times 10^6$ \msol\ black 
hole (Ghez et al. 2005; Gillessen et al. 2009; Reid and Brunthaler 2004). The extremely high spatial resolution 
made possible by its relative proximity provides the best laboratory for studying the properties of 
low-luminosity accreting black holes; 1$''$ corresponds to 0.039 pc at the Galactic center distance of 8 kpc 
(Reid 1993). The emission from Sgr~A* is assumed to be produced from radiatively inefficient accretion flow, 
as well as outflows. The bulk of the continuum flux from Sgr~A* is considered to be generated in an accretion 
disk, where identifying the source of variable continuum emission becomes essential for our understanding of 
the launching and transport of energy in the nuclei of galaxies.

The emission from Sgr~A* consists of both quiescent and variable components. The strongest variable component is 
detected as flares at near-IR and X-ray wavelengths (Baganoff et al. 2003; Genzel et al. 2003; 
Goldwurm et al. 2003; Eckart et al. 2006; Yusef-Zadeh et al. 2006a; Hornstein et al. 2007; Porquet et al. 2008; 
Dodds-Eden et al. 2009; Sabha et al. 2010; Trap et al. 2011), whereas only moderate flux variation is found at 
radio and sub-mm wavelengths (Falcke et al. 1998; Zhao et al. 2001; Herrnstein et al. 2004; Miyazaki et al. 2004;
Yusef-Zadeh et al. 2006b; Marrone et al. 2008).  
The spectral energy distribution (SED) of the quiescent component peaks at sub-mm wavelengths and is 
identified in radio, millimeter and sub-mm wavelengths (see Genzel et al. 2010 and references therein). This 
emission is thought to be produced by synchrotron radiation from a thermal population of electrons with 
$kT\sim 10-30$\,MeV participating in an accretion flow. A variety of models have been proposed to explain the 
quiescent emission from Sgr~A* by fitting its SED, including a thin accretion disk, a disk and jet, an 
outflow, an advection-dominated accretion flow, a radiatively inefficient accretion flow, and advection-dominated 
inflow/outflow solutions (Blandford \& Begelman 1999; Melia \& Falcke 2001; Yuan et al. 2003; Liu et al.2004; 
Genzel et al. 2011 and references cited therein).  
Unlike the quiescent component, which originates over a wide range of physical conditions and length scales 
of the accretion flow, flares are localized, allowing emission models to be directly tested with observations.  
As a supermassive black hole candidate, Sgr~A* presents an unparalleled opportunity to 
closely study the process by which gas is captured, accreted, or ejected, by characterizing the emission 
variability over timescales of minutes to months. Because the time scale for variability is proportional to the 
mass of the black hole, this corresponds to variability on timescales 100 times longer than that of more 
massive black holes in the nuclei of other galaxies.

Studying near-IR emission from Sgr~A* is crucial to track the acceleration of energetic particles, as well as the 
accretion flow.  Near-IR flares are produced by synchrotron radiation from a transient population of accelerated 
electrons.  The near-IR emission is dominated by flaring activity that occurs a few times per day, with a small 
fraction of events showing simultaneous X-ray flares.  The X-ray flare mechanism has been interpreted as either 
inverse Compton scattering (ICS), self-synchrotron-Compton (SSC), or synchrotron emission (Markoff et al. 2001;
Liu \& Melia 2002;  Yuan et al. 2004; Yusef-Zadeh et al. 2006a; Eckart et al. 2009; Marrone et al. 2008; 
Dodds-Eden et al. 2009). The X-ray synchrotron mechanism implies that the acceleration mechanism 
must continuously resupply the 100\,GeV electrons for the 30 minute duration of the observed flares, because
the synchrotron loss time of the $\sim 100$\,GeV electrons that are responsible for the synchrotron emission 
is $\sim 30$\,seconds. The synchrotron self-Compton model requires that the local magnetic field be extremely 
large or that the number density of electrons is high. This is necessary to avoid overproducing the near-IR 
synchrotron emission from the large number of energetic electrons that are required to upscatter infrared 
photons into the X-ray band (Dodds-Eden et al. 2009;  Marrone et al. 2008; Sabha et al. 2010; Trap et al 2011). 
The typical parameters of the magnetic field --- B$\sim$ 1-10 G or electron density n$_e\sim 10^9$ cm$^{-3}$ ---
correspond to an energy density in the accelerated electrons a thousand times larger than that in the magnetic 
field.  It is then difficult to understand how these particles are accelerated and confined. 

In the case of X-ray emission produced by inverse Compton scattering, two possibilities have been explored. 
First, sub-mm photons arising from the quiescent component of Sgr A* may be upscattered by the transient electron 
population that is producing the IR synchrotron emission during IR flares (Yusef-Zadeh et al. 2006a). 
Alternatively, near-IR photons emitted during the flare may be upscattered by the mildly relativistic 
$\sim 20$\,MeV electrons responsible for the quiescent radio--submm emission 
(Yusef-Zadeh at al.\ 2006a, 2008, 2009).  If the sub-mm emission region were optically thin, this would 
produce a similar X-ray luminosity as the upscattering of sub-mm seed photons.  However, because the sub-mm source 
is optically thick below $\sim$1000\,GHz, the observed sub-mm flux is produced by a fraction of the underlying 
electrons.  The exact frequency at which the quiescent emission becomes optically thick is unknown. 
However, sub-mm measurements between 230 and 690 GHz (Marrone et al.  2006) indicate a flattening of the spectral 
index and thus a deviation from the rising spectrum observed at lower frequencies (An et al. 2005). 
The emission region is optically thin to near-IR photons, so all of these electrons are available to 
upscatter near-IR seed photons to X-ray energies (Yusef-Zadeh et al. 2009).  
The ICS luminosity produced through this scenario compares favorably with the observed near-IR and X-ray 
luminosities (Yusef-Zadeh et al. 2009). This is the model on which we will focus, as described below.

One of the predictions of the ICS model, in which near-IR photons are upscattered by $\sim 10-30$\,MeV electrons,
is a time delay between the peaks of the near-IR and X-ray flares (Yusef-Zadeh et al. 2009; Dodds-Eden et al. 2009). 
Wardle (2011) provided the theoretical framework for the X-ray echo picture of the ICS. 
We present evidence for a time delay between the peaks of X-ray and near-IR flare emission  
based on seven new and archival observations. 
These measurements provide support for X-ray  production via inverse-Compton scattering of IR flare 
photons by relativistic electrons of the accretion flow. 
The cross-correlation profiles of the peaks are generally skewed toward positive time lags,
but show maximum likelihood values that have low signal-to-noise, due to the limited number of detections
of simultaneous X-ray and near-IR flares.

\section{Observations}

\subsection{X-rays}

X-ray observations used in this study come from the Chandra observatory.
Data obtained on July 6--7, 2004 and July 30, 2005 consist of 50.2 and 46 kilo-second (ks)
observations, respectively,  (ObsIDs 4683,5953), which were described previously 
by Eckart et al. (2006) and Muno et al. (2005).  Data obtained in 2008 (not previously 
reported) consist of six 28 ks observations, starting May 5, May 6, May 10, May 11, 
July 25, and July 26 (ObsIDs 9169, 9170, 9171, 9172, 9174, 9173 respectively),
scheduled to match nighttime IR observations in Chile (see below).  
All observations placed Sgr A* at the ACIS-I aimpoint and took data in FAINT mode.

We checked for any time intervals of strong background flaring (none were found) and then reprocessed the data using
CIAO 4.3\footnote{e.g. http://cxc.harvard.edu/ciao/threads/createL2/}. This involved applying corrections to the
energy scales to compensate for time-dependent gain changes and charge-transfer inefficiency, removing pixel
randomization and improving spatial resolution, as well as creating an updated bad pixel map. We filtered the data for
"bad" grades and status.

We extracted lightcurves (in spacecraft TT time) from a 1$''$ circular region around the position of Sgr A*, in the
1.5-8 keV energy range, using Gehrels (1986) errors. We converted the timestamps to UTC time following the
prescription by
A. Rots\footnote{http://cxc.harvard.edu/contrib/arots/time\_tutorial.html/}. 
The baseline quiescent
X-ray emission from Sgr A* is spatially extended (Baganoff et al. 2003), but we see no variations in other local background
emission. We tested several choices of binning the data for comparison to other wavelengths, settling on 1500 s
binning for the 2008 data, 300 s binning for the major flare on 6--7 July 2004, and 600 s binning for the flare on
July 30, 2005. Using the absorbed thermal plasma model of Baganoff et al., the ratio of 1.5-8 keV counts to 2-10
keV unabsorbed flux is $8\times10^{-11}$ ergs cm$^{-2}$ s$^{-1}$ per count/s. 
Recent measurements indicate a distance of 8.3 kpc to Sgr A* (Gillessen et al. 2009),
but we assume a distance of 8 kpc,
which gives $L_X$(2-10)$=6\times10^{35}$ ergs/s per count/s, or a typical quiescent luminosity of $3\times10^{33}$
ergs/s, in agreement with Baganoff et al. (2003).

We used Kolmogorov-Smirnov (K--S) tests
(using the lcstats FTOOL \footnote{http://heasarc.gsfc.nasa.gov/lheasoft/ftools/xronos.html}) on
the 2008 Chandra lightcurves (binned to 32.41 s) to search for evidence of variability.  We find evidence for
variability in three of the 2008 observations, while another three show no evidence of variability. ObsIDs 9169,
9172, and 9173 give K--S probabilities of a constant lightcurve of $5\times10^{-6}$, $2\times10^{-4}$, and
$1\times10^{-8}$, respectively, while the remaining observations give K--S probabilities greater than 5\%. 
This significantly strengthens the evidence of variability from Sgr A* at very low levels, as Baganoff et al.
(2003) reported a much larger K--S probability of constancy of $7\times10^{-3}$ during quiescence.
Given that the quiescent X-ray emission arises from much larger scales, 
presumably due to Bondi-Hoyle accretion (Baganoff et al. 2003), we suggest that the  X-ray variability noted 
here is due to low-level flare emission  superimposed on the steady quiescent emission. 
Alternatively, the X-ray variability on hourly time scales could arise 
from  coronally active stars producing giant  flares (Sazonov, Sunyaev and Revnivtsev 2011).

\subsection{Near-IR}

For the near-IR observations we use archival data taken with the VLT and HST. 
The near-IR data taken in 2004, 2005 and 2008 were observed with the near-IR adaptive
optics-assisted imager NACO at the VLT (Lenzen et al. 2003; Rousset et al. 2003). 
We used Ks-band 13-milliarcsecond (mas) pixel
imaging data from 2004 (nights of July 6-7) and 2005 (night of July 30-31) first presented by 
Eckart  et al. (2006, 2008),  and   Ks-band 13-mas pixel polarimetric imaging data from
2008 (nights of May 4/5, 5/6, 9/10, 10/11 and July 25/26, 26/27) first presented
in Dodds-Eden  et al. (2011). 
We did not apply any data quality cut for the latter observations, except for the
elimination of 11 images from July 25th, 2008, due to a bad AO correction
(quadfoils, or a 'waffle' pattern).

We used both aperture photometry and PSF photometry methods to produce light curves,
both carried out in the way described in Dodds-Eden  et al. (2011). 
In particular we note that, for the purposes of that paper, the aperture method
used two small apertures, one centered on the position of Sgr A* and the other on
the star S17 (confused with Sgr A* in 2006--2008), in order to measure their combined flux.
Since S17 was further from Sgr A* in 2004 and the combined measurement of the flux
is not important for the purposes of this paper, the use of the above method
unnecessarily decreases the S/N. As a result, for the near-IR/X-ray comparison we
supplemented the dataset with higher S/N light curves obtained from PSF photometry,
though the data were sparsely sampled. 
We provide the light curve of S17 in the 2008 data set and the light curve of the comparison star S7 for 
the 2004 and 2005 data sets. 
The stellar background is estimated to be $3.4 \pm 0.2$ mJy in 2008 (Dodds-Eden et al. 2011). 


Near-IR HST observations used in this study are NICMOS archival data
obtained on 4 April 2007 as part of a larger Sgr A* monitoring campaign.
Full observational details have been presented in Yusef-Zadeh et al.
(2009). Briefly, the exposures used NICMOS camera 1, which has a pixel
scale of 0.043$"$, and the medium-band filters F170M and F145M, which have central
wavelengths of 1.71 and 1.45 $\mu$m, respectively, and FWHMs of 0.2 $\mu$m. Individual exposures
had a duration of 144 sec, with non-destructive detector readouts occurring
every 16 sec. We averaged the readouts to sampling intervals of 64 and 128 seconds in
the 1.71 and 1.45 $\mu$m bands, respectively, to obtain adequate S/N. 
Aperture photometry was performed on Sgr A* in each sampling interval, using an aperture diameter of 3 pixels in
order to limit contamination by nearby stars. 

All of the near-IR measurements presented here have been corrected
for reddening using extinction values of A$_\lambda$ = 2.42 (2.2$\mu$m), 4.34 (1.71$\mu$m),
and 6.07 (1.45$\mu$m) from Fritz et al. (2011).


\section{Results}

The middle two panels of Figure 1 show
the light curves from the archival Ks-band and X-ray observations of Sgr A* 
that were taken on 2005 July 30. 
The near-IR light curve of the comparison star S7 is shown in the top panel.  
The X-ray and near-IR light curves of Sgr A* were sampled at intervals of 200 and 600 s, respectively.
These measurements, first reported in Eckart et al. (2008), indicate a flare with a peak X-ray 
luminosity of 8$\times10^{33}$ erg s$^{-1}$. The 
bottom panel shows the cross-correlation of these light curves. 
The cross-correlation analysis uses the Z-transformed 
discrete correlation function (ZDCF) algorithm (Alexander 1997). The ZDC function is an improved solution to the 
problem of investigating correlation in unevenly sampled light curves. 
Maximum likelihood values, as well as 1 and 2$\sigma$ confidence intervals around those values are estimated  
using the start time of each bin. This analysis finds that the X-ray 
peak lags the near-IR peak in Figure 1 by $\sim$8.0 (+10, -10.1) and 8.0 (+20.2, -17.9) minutes for 1 and 2$\sigma$ 
maximum likelihood values, respectively. 
We varied the sampling interval in the 
near-IR and X-ray data between 1.5  and 10 minutes, but the maximum likelihood lag value remained the 
same. Eckart et al. (2008) compared the X-ray and near-IR flare emission and found that the peaks are 
coincident within $\pm7$ minutes.


The aperture photometry technique that was used to reanalyze the near-IR VLT data from 2004
produced a light curve that is quite similar that published by Eckart et al. (2006), who 
deconvolved their images. The only difference is that the present analysis uses data extending
up to 4h UT on July 7, which is longer than that of Eckart et al. (2006).  
The bottom three panels of Figure 2 show the cross-correlation of these near-IR and X-ray data 
with a maximum likelihood lag of 7.0 (+1.3, -1.1) minutes and (+7.5, -6.9) minutes with 1 and 
2$\sigma$ errors, respectively. 
The lag is larger than zero at the 1$\sigma$ level.  The near-IR light curve of the 
comparison star S7 is shown in the top panel.  The light curve of S7 is constant and supports 
the variable emission from Sgr A* between 3 and 4h UT. The sampling of the near-IR data reduced using
PSF photometry is much more sparse than the aperture photometry data.
The 1$\sigma$ cross-correlation peak using the PSF photometry showed a time lag 7.7 (+2.9, -3.4) minutes 
within the  error bars of that reduced by aperture photometry. Eckart et al. (2006) also 
cross-correlated their data and showed no time delay within 10 minutes.


Next we examined the X-ray and near-IR light curves obtained on 
2007 April 4 using XMM and VLT observations (flare \#2 in Porquet et al.\ 2008; 
Dodds-Eden et al.\ 2009). These data contain the second 
brightest X-ray flare (flare \#2 in Porquet et al;\ 2008) coincident with one of the 
strongest near-IR flares, that has ever been detected.  
The cross-correlation of the X-ray and near-IR data for this bright flare
shows a 1$\sigma$  maximum likelihood time delay of -0.5 (+7.0, -6.5) minutes, which
is consistent with zero time delay (Dodds-Eden et al.\ 2009; Yusef-Zadeh et al.\ 2009). 
The peak luminosity of the brightest flare is 24.6$\times10^{34}$ erg s$^{-1}$ (Porquet et al.\ 2008). 
Two other moderate X-ray flares (flares \#4-5) were detected 
on 2004 April 4 following the bright X-ray flare. 
X-ray flares \#4 and \#5, with peak luminosities of $\sim6\times10^{34}$ and  
$\sim8.9\times10^{34}$ erg s$^{-1}$, respectively, are covered in the near-IR 
1.71 and 1.45$\mu$m NICMOS data. 
The cross-correlations of the X-ray and near-IR light curves for flares \#4-5 are presented in Figure 3a-d.
The NICMOS observations alternated between the 1.71 and 1.45$\mu$m bands every 6 minutes. 
In all of the four cases studied, the maximum likelihood values of flares \#4 and \#5
show positive lags ranging between 5 and 10 minutes.  
Similar to the other cases  analyzed here, the peaks of the cross-correlations are all  
skewed towards positive time lags.

Finally, we compared near-IR (VLT) and X-ray data taken in May and July 2008. Figure 
4 shows the light curves from the two  different days of observations. 
These X-ray flares are an 
order of magnitude less luminous than those detected in earlier observations. We have carried out K--S 
tests indicating the reality of these low-level X-ray visibilities (\S 2.1).  
The cross-correlations of the light curves from these two days give maximum likelihood lags of 
19 (+6.8, -2.4) and 14.6 (5.6, -2.9) minutes with 2$\sigma$ error bars.  
Figure 4 also shows the cross-correlation of X-ray with near-IR light curves 
derived from PSF photometry. 
The resulting time delays of 26.5 (+19, -29) and 16.6 (14.8, -11.8) minutes 
are well within the error bars of the aperture photometry  data.  
To provide additional 
support for the reality of the variability of Sgr A*, Figure 5 compares
the  light curves of Sgr A* and S17 using the 2008 May 5  and 2008 July 26 observations,
which are based on PSF photometry.
In these data, where Sgr A* and S17 are separated from each other, each source
is detected independently.
  
Although most of the individual cross-correlation results that are presented 
here have low S/N, the 1$\sigma$ maximum likelihood peaks in eight different measurements 
show a tendency for X-ray emission to lag near-IR emission rather than lead.  
The cross-correlation gives maximum likelihood near-IR-to-X-ray lag values that are 
systematically higher than zero. The strongest simultaneous near-IR and X-ray flares 
(Flare 2 in Porquet et al. 2008) do not show any time delay, whereas the 
faintest X-ray flares seem to show the longest time delays.

\section{SSC Models}

Several alternative models for the relationship between near-IR and X-ray 
flares have been proposed.  Synchrotron emission from a 
high-energy tail of the accelerated electron population responsible for the 
near-IR flaring may be responsible for the observed X-ray flaring 
(Dodds-Eden et al.\ 2009, 2010).
Self-synchrotron-Compton (SSC) models, in which the same population of electrons 
produce the near-IR synchrotron emission and upscatter lower-energy synchrotron 
photons, require an unrealistically compact, and hence 
over-pressured, source region (Dodds-Eden et al.\ 2009) or a very weak 
magnetic field or a high electron density to avoid overproducing the 
IR synchrotron emission (Marrone et al. 2008; Sabha et al. 2010; Trap et al. 2011). 

In SSC models the observed ratio of the X-ray and IR fluxes
demands a certain Thomson optical depth in relativistic electrons.  SSC flare
models (Marrone et al 2008; Sabha et al. 2010; Trap et al 2011) adopt source
region radii R of order R$_s$. The requisite electron densities then imply a
particular magnetic field strength, so that the synchrotron emission from the
electron population matches the observed near-IR flaring.
These SSC models have electron energy densities ranging between 10$^3$ and 2$\times10^5$
times the magnetic energy density.  Because electron acceleration mechanisms invoke
magnetic fields, the energy density in the field should be comparable to or
greater than the energy density of the accelerated particles.  Thus,  scaling the
SSC models to equipartition fields by reducing the electron density while
increasing the magnetic field to keep the product n$_e \rm B^{[(p+1)/2]}$ and hence the
near-IR synchrotron flux fixed, one finds that a reduction in electron density by a
factor of 40 or more is required.  Thus, SSC contributes at most 1/40 of the
observed X-ray flux in such a model.

Alternatively, one can attempt to construct SSC models in which the field is in
equipartition by reducing the source size R, while keeping the products
n$_e \rm B^{[(p+1)/2]} \rm R^3$ and n$_e \rm R$ fixed to preserve the synchrotron and SSC fluxes, 
respectively.  Equipartition between the relativistic electrons and magnetic
field is attained when R is reduced by a factor of a thousand or more, with
corresponding field strength in the 10$^4 - 10^5$ G range, which is orders of magnitude
more than what is reasonable.

\section{X-ray Echo due to ICS}

  
Here we focus on an inverse Compton scenario for the X-ray flares, 
suggested by Yusef-Zadeh et al.\ (2009) and outlined in more detail by 
Wardle (2011).  In this model, near-IR flare photons are upscattered to X-ray 
energies by the thermal electrons ($kT_e\sim 10$\,MeV) in the accretion 
flow.  This process dominates the alternative inverse Compton pathway, in 
which the nonthermal energetic electrons responsible for the near-IR 
synchrotron emission upscatter sub-millimeter photons emitted by the 
thermal electrons in the accretion flow into the X-ray band 
(Rybicki \& Lightman  1986, Chapter 7.5). 
This alternative ICS pathway is less effective
because the accretion flow is optically thick in the 
sub-millimeter, so that the ratio between sub-mm photons and the thermal 
electrons producing them is reduced by a factor of the optical depth.  
Then the upscattering of near-IR photons proportionately produces more 
emission 
than would be inferred by implicitly assuming that the sub-millimeter 
synchrotron flux is optically thin (e.g.\ Dodds-Eden et al. 2009).
In this process, the second order scattering echo  can also 
produce  MeV $\gamma$-ray emission with a  
luminosity L$_{\gamma}$ that is lower than that of X-rays L$_X$ 
 by a factor of a few (ie L$_{\gamma}$/L$_X \sim$ L$_X$/L$_{NIR}$).

One significant difference of the ICS picture from  the synchrotron and SSC pictures 
is that the longer path from the 
near-IR source to the observer taken by an upscattered photon detected in the X-ray 
compared to the straight-line path taken by a photon received in the near-IR introduces a 
time delay between flaring in the near-IR and X-rays.  In addition, because scattering occurs 
from a range of locations within the accretion flow, with a corresponding range of time 
delays, the reflection signal tends to be broadened compared to the near-IR seed photon light 
curve.  While there is some evidence of systematic delays between the near-IR and X-ray flaring, 
the X-ray flares appear to generally have a narrow FWHM compared to their corresponding near-IR 
flares.

\subsection{Modeling}

To explore whether this model can plausibly explain the X-ray flaring, we compute the X-ray 
``echoes'' of the observed near-IR flares to compare with our simultaneous X-ray 
observations.  We make a number of simplifying assumptions, none of which are severe.  We 
assume that the observed near-IR flare comes from a point located in the accretion flow 
with a power-law spectrum and Gaussian light curve, 
$S_\nu(t) \propto \nu^{-0.5} \exp(-(t-t_0)^2/2\sigma^2)$. 
Because the X-ray flares are narrower than the near-IR we have 
assumed that the FWHM of the near-IR flaring narrows  
as $\lambda^{0.5}$ below 2.2\micron.  The physical justification is that the 
synchrotron loss time scale scales as $\lambda^{0.5}$ and becomes comparable with the 
observed 
FWHM at about 2\micron.  The energy of an upscattered photon with initial energy 
$h\nu_\mathrm{IR}$ is assumed to be $\frac{4}{3}\gamma^2 h\nu_\mathrm{IR}$ where $\gamma$ 
is the electron Lorentz factor.  Because the upscattered photon energies are much lower than the 
electron rest energy,  the scattering occurs in the Thompson regime. 
Assuming isotropic 
upscattering, the total production rate of upscattered photons per unit volume is 
$n_\mathrm{IR} n_e\sigma_T c$, where $n_\mathrm{IR}$ and $n_e$ are the number densities of 
infrared photons and relativistic electrons, respectively.  We ignore relativistic effects such as the 
Doppler boosting associated with the bulk motion of the accretion flow, because the 
corresponding Lorentz factor is small compared to that of the individual electrons.  We 
also ignore the time delay, gravitational redshift and lensing effects of the Kerr metric,
which only become important close to the event horizon in highly inclined systems.

The electron density and temperature profiles in the accretion flow are assumed to be steady, axisymmetric power-laws 
in cylindrical radius $r$, with $n_e \propto r^{-0.75}$ and $T_e \propto r^{-1}$ and the density truncated within 
$2\,R_s$ and beyond $20\,R_s$.  The accretion flow is assumed to be confined to a thick disk with scale height/radius 
($h/r$) = 0.5. The electron population is characterized by an approximate relativistic Maxwellian $f(x) = 1/2 \, 
x^2\exp(-x)$, where $x = E / (kT_e)$, which is a good approximation for $kT_e \ga 2$\,MeV.
The adopted profiles are within the 
typical ranges considered in analytic estimates (e.g.\ Loeb \& Waxman 2007), semi-analytic models for the accretion 
flow (e.g. Yuan, Quataert \& Narayan 2003), and MHD simulations (e.g.\ Mo\'scibrodzka et al.\ 2009).

The remaining parameters specify the flare location relative to the line of sight and relative to the accretion flow: 
the inclination $i$ of the disk to the line of sight, and the flare location $(r,\phi,z)$ in the natural cylindrical 
coordinate system. The low optical depth of the accretion flow to near-IR photons means that the results are 
insensitive to the inclination $i$ and the azimuthal angle $\phi$ between the flare location and the poloidal plane 
containing the line of sight and the z-axis. We therefore fix these at typical values $i=45^{\circ}$ and $\phi=90^\circ$, 
respectively.  Similarly, the results are insensitive to the height $z$ of the near-IR flare for $z\la r$, so we 
simply assume that the flare occurs in the disk midplane, ie.\ that $z=0$.

The free parameters are the electron density $n_0$ and temperature $T_0$ at the fiducial radius $2\times10^{12}$\,cm, 
and the radial location of the flare, $r$. The noisiness of the observed light curves preclude formal fitting, so for 
each near-IR/X-ray flare combination we adjust these parameters by hand to approximately match the X-ray light curve. 
Reasonable matches to the observed X-ray profile are obtained with flares occurring at $r\sim10 R_s$, and electron 
densities $\sim 10^{7.5}$--$10^{8.5}$\,cm$^{-3}$ and temperatures $\sim 5$--20\,MeV, as listed in Table 1.

\begin{deluxetable}{lrrr}
\tablewidth{0pt}
\tablecaption{Model Parameters for Flare Events}
\tablehead{
\colhead{Flare}&
\colhead{$kT_0^a$ (MeV)}&
\colhead{$n_0^b$ (cm$^{-3})$}&
\colhead{$r^c$ (cm)}
}
\startdata
2004 \,Jul 07  &  9   & $2.9\times10^8$ & $1.0\times10^{13}$ \\
2005 \,Jul 30  & 20   & $2.3\times10^7$ & $1.0\times10^{13}$ \\
2007 Apr 04 \#2   &  7   & $2.1\times10^8$ & $4.0\times10^{12}$ \\
2007 Apr 04 \#4 &  5   & $7.9\times10^8$ & $1.0\times10^{13}$ \\
\enddata
\label{tab:fitparams}

\tablenotetext{a} {Accretion  flow temperature profile $T_e(r)=T_0(r/r_0)^{-1}$ where $r_0=2\times10^{12}$ cm}

\tablenotetext{b} {Density profile $n_e(r,z)=n_0 (r/r_0)^{0.75}\rm {exp}(-2z^2/r^2)$}

\tablenotetext{c} {Radial location of near-IR flare in equatorial plane. (R$_s \approx 1.2\times10^{12}$ cm for 
M$_{\rm BH} = 4\times10^6 $ \msol)}

\end{deluxetable}

In the context of the ICS picture, we fit a sample of light curves that have good time coverage in near-IR and X-ray 
wavelengths in order to illustrate the point that this model can potentially be a powerful tool to quantify the 
physical characteristics of the accretion flow. The light curves presented in Figures 1 and 2 are modeled following 
the second brightest X-ray flare that has ever been recorded on 2007 April 4 (Porquet et al. 2008).  The light curves 
of the moderate flare that followed this bright flare (flare \#2) is presented in Figure 3. The time delays of the 
peak emission shown in Figures 1, 2 and 3 are 8, 7 and 8 minutes, respectively, whereas the bright flare on 2007 
April 4 showed a time delay consistent with zero. Given the limited simultaneous time coverage of the flares 
shown in Figure 3 and 4, we focus only on modeling these four flares.  Figure 6 shows the observed and modeled 
light curves for the simultaneous near-IR and X-ray flares that occurred on 2005 July 30, 2004 July 
6/7 and 2007 April 04 (the main flare and flare \#4). 
Parameters of the fit for each of the four examples are shown in Table 1. 
Substructures corresponding to two weak flares in Figure 6a are 
also modeled with the same parameters as the main flare, as listed in Table 1.   
A baseline level has been subtracted from the near-IR light curves before constructing the theoretical light curves.
If we restrict 
the evaluation of $\chi^2$ to just the flare part of the X-ray light curve, $\chi^2$/df is between 3--4. The reason is 
that there is often point-to-point variability in the light curve that throws points well away from the smooth 
$''$prediction$''$. 
In addition, the near-IR light curve in Figure 6c shows substructures that 
could be arising  from different flares, which would have   different time delays in the 
context of ICS. Thus we have not formally fit the light 
curves as our models 
are illustrative only.  Given these limitations, we obtain reasonable parameters from 
the theoretical X-ray light curves,
which are superimposed on the observed light curves in Figure 6.


In each case a Gaussian form of the near-IR light
      curve has been adopted to represent the observed
      (extinction-corrected) near-IR flare at 3.8\micron,
      2.2\micron\ and 1.7\micron. However, in our models,
      the X-ray flare arises from scattered optical
      photons. We assume the peak flux of the synchrotron
      flare (emitting at near-IR to optical) scales as
      $\nu^{-0.5}$ with frequency.  Because the X-ray flare
      is narrower than the near-IR we have assumed
      that the FWHM is constant below 2.2\micron\ and
      narrows as $\lambda^{0.5}$ shortward of this.
      Reasonable matches to the observed X-ray profile are
      obtained with the flare occurring at the inner edge
      of the density profile and electron densities, as
      their estimated values are given in Table 1. The FWHM
      assumption requires both the rise and the decay
      timescale of the flare to be faster at optical
      frequencies than at near-IR. For the decay timescale,
      this is a natural result of synchrotron cooling (the
      synchrotron loss time scale scales as $\nu^{-0.5}$).
      The rise time scale depends on the acceleration
      mechanism. While the acceleration mechanism of flare
      production is not understood, it is plausible that
      optically-emitting electrons are produced later than
      IR-emitting electrons (Kusunose and Takahara 2011). 


\begin{deluxetable}{lrrrrrr}
\tablewidth{0pt}
\tablecaption{Flux Ratios vs. Time Delay}
\tablehead{
\colhead{Flare}&
\colhead{IR}&
\colhead{X-ray} &
\colhead{IR peak}&
\colhead{X-ray Peak}&
\colhead{Peak Ratio}&
\colhead{Time Delay}\\
\colhead{}&
\colhead{backg.}&
\colhead{backg.}&
\colhead{mJy}&
\colhead{1$\times10^{35}$ erg s$^{-1}$}&
\colhead{IR/X-ray}&
\colhead{(minutes 1$\sigma$}
}
\startdata
 2007 Apr 04 \#2   &  5.0   & 0.2 & 16.5  & 4.8   &  3.4    & -0.5  (+7, -6.5) \\
 2007 Apr 04 \#5 &  -0.2   & 0.2 &  2.63 & 1.23   & 2.1 &  5.0 (+1.9, -1.5)  \\
 2007 Apr 04 \#4 &  -0.2   & 0.2 &  4.67 & 1.22   & 3.8 &  5.0 (+1, -1.4)  \\
 2004 \,Jul 07  &  1.5   & 0.03   &  6  & 0.39  & 15.5 &  7 (+1.3 -1.2) \\
 2005 \,Jul 30  &   0.0    & 0.002   & 5.9  & 0.13 & 45.4  &  8 (+10, -10.1)  \\
 2008 \,July 26+27  &  5.5    & 0.003   & 3.7  & 0.05 & 68.5  &  14.6 (+5.6, -7.4)  \\
 2008 \,May 5  &   6    & 0.003   & 2.74  & 0.021 & 130.0  &  19 (+6.8, -2.4)  \\
\enddata
\label{tab:fitparams}
\end{deluxetable}

Another issue involves the prediction that the
      spectral index between X-rays and near-IR/optical
      emission be identical in the context of ICS with the
      assumption that the electron distribution has a
      single power law spectrum.  It is, however, possible
      that the energy spectrum of electrons has a broken
      power-law, thus producing a different spectral index in
      near-IR and X-rays. The mismatch in spectral index is
      in fact noted for the bright X-ray flare coincident
      with a strong near-IR flare of 2007 April 4
      (Dodds-Eden et al. 2009). Although these authors
      discuss the difference in the spectral index using
      a synchrotron mechanism, the broken power law of NIR
      emitting electrons with a steeper spectral index
      shortward of 3.8$\mu$m was also argued in the ICS
      scenario (Yusef-Zadeh et al. 2009).

      Although the spectral index measurements in X-ray and
      near-IR can not distinguish between the synchrotron
      and ICS models for the production of X-rays, it is 
predicted  that the ratio of near-IR to X-ray flare emission  
can increase  with increasing time delay in the ICS scenario.  
This is because bright X-ray flares are generated in the inner disk 
where the 
time delay is expected to be small. Although the available data are limited to test this aspect of the proposed model, 
we note a trend that is consistent with this expectation. Table 2 shows the ratio of 2.2 $\mu$m  peak flux (mJy) to 
peak X-ray luminosity (10$^{35}$ erg s$^{-1}$) for seven   different measurements. 
The 1$\sigma$ error bars of the maximum likelihood values are given in column 7. 
The  smallest to largest flux ratios, as shown in column 6 of Table 2, 
support the trend that near-IR/X-ray flux ratios increase with the time delay, as 
expected in the context of  the ICS model. 
For the near-IR  observations on 2007 April 4 obtained at 3.8 and 1.7$\mu$m, 
we convert the peak flux to 
2.2$\mu$m  using the spectral index of -0.7,  where S$_{\nu}\propto \nu^{-0.7}$, 
before we estimate the flux ratio.   
Future simultaneous measurements of X-ray and near-IR flares should examine the correlation of the 
peak near-IR to X-ray flux as a function of increasing observed time delay.

In summary, we have presented cross-correlations of simultaneous X-ray and near-IR flare light curves from Sgr A* and 
found a time lag of the X-ray peak flare emission with respect to the near-IR.  Such an X-ray echo provides support 
for inverse Compton scattering of near-IR flare photons by $\sim 5-20$\,MeV electrons. 
A fraction of near-IR flare photons must upscatter from the accretion flow into the X-ray band.  This can be 
significant for plausible accretion models, and therefore may explain the observed X-ray flares, or at least place significant 
constraints on the accretion flow.
Future cross-correlations based on 
more continuous near-IR and X-ray observations should give us better S/N in the maximum likelihood values of the time lag.
In the context of the ICS model, future measurements will place better constraints on the density and temperature 
profiles of the accretion flow and the location of near-IR flares.

\acknowledgments
This work is partially supported by grants AST-0807400 from the National Science Foundation and DP0986386 from the Australian Research Council. COH is supported by NSERC and an Ingenuity New Faculty Award.

\begin{figure}
\center
\includegraphics[scale=0.4,angle=0]{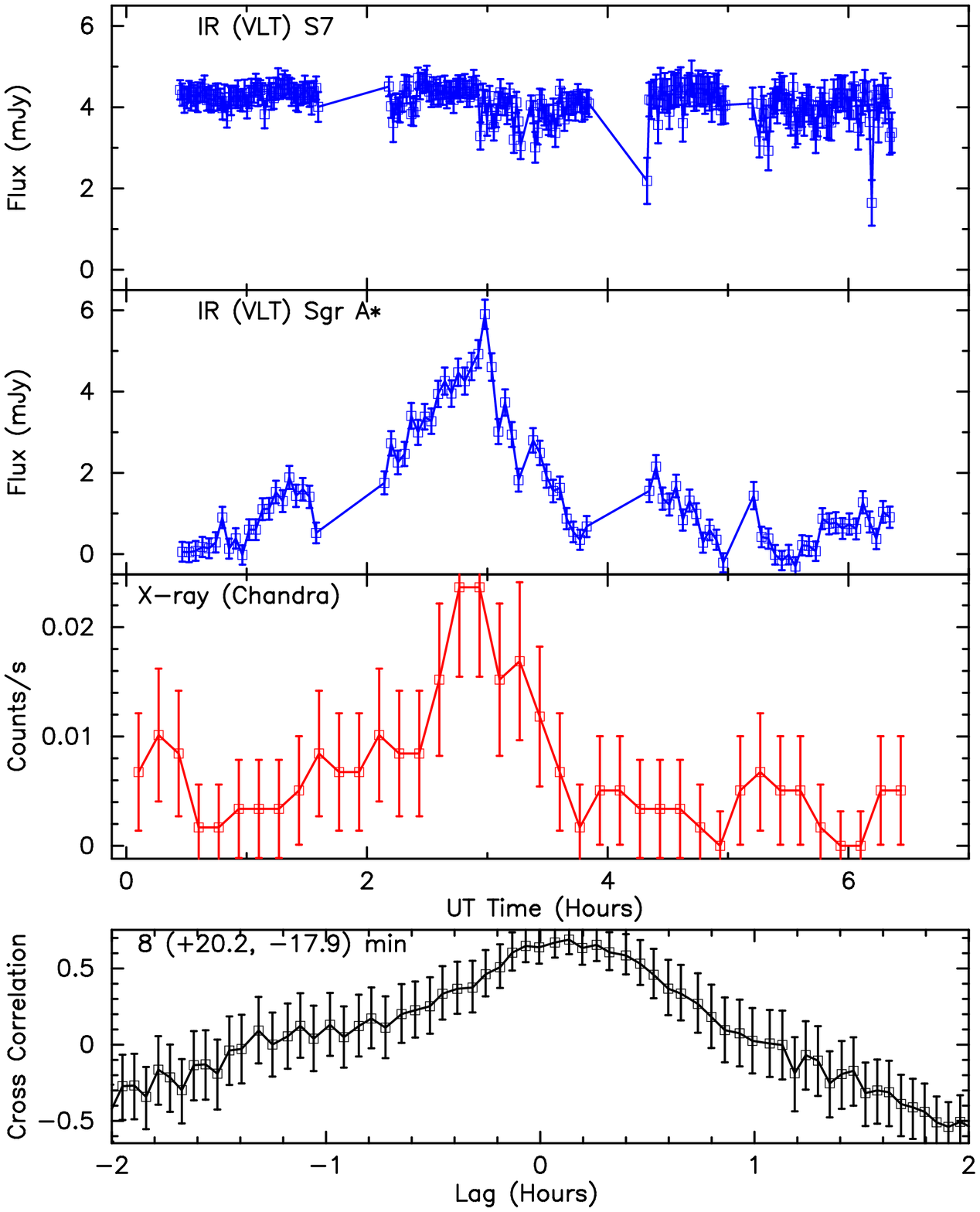}
\caption{
The light curve of the comparison star S7 is shown in the top panel with 
a time sampling 
of 65 sec. 
Two middle panels show the light curves of Ks-band (2.2$\mu$m)  
and X-ray (2-8keV) data taken simultaneously by the VLT and 
Chandra on 2005 July 30 (Eckart et al. 2008). 
The time sampling for the X-ray and near-IR data are 600 and 200 seconds, 
respectively. The cross correlation of the lightcurves and the corresponding 2$\sigma$
 maximum 
likelihood values  are shown in the bottom panel.
The 1$\sigma$ error bar is given  in Table 2. 
A base level of 4.2 mJy  have been  subtracted
from the lightcurve of Sgr A* (Dodds-Eden et al. 2011). 
} 
\end{figure}

\begin{figure}
\center
\includegraphics[scale=0.4,angle=0]{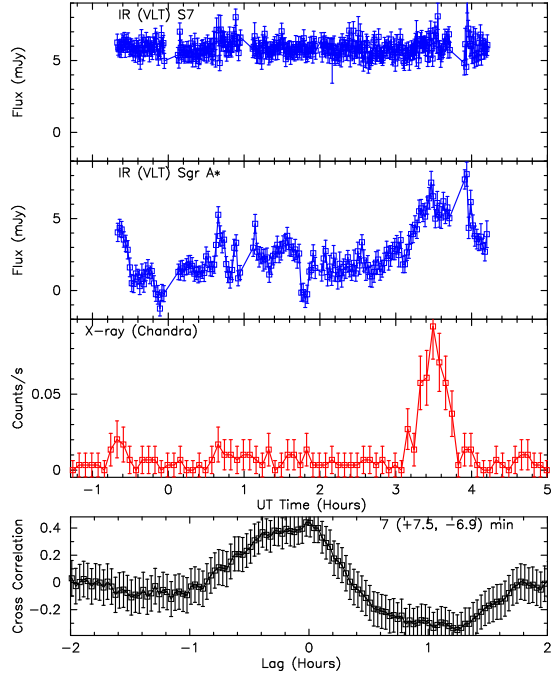}
\caption{
The top panel shows K-band (2.2$\mu$m) VLT data of the comparison star 
on S7 with a time sampling of 45 seconds and Sgr A* on 2004 July 6/7, while the third panel shows simultaneous 
Chandra X-ray (2-8
keV) data (Eckart et al. 2006).
The   middle  two panels show the light curves of K-band (2.2$\mu$m)  
and X-ray (2-8 keV) data taken simultaneously by the VLT and 
Chandra on 2004 July 6/7 (Eckart et al. 2006). 
The time sampling for the X-ray and near-IR data on Sgr A* are 300 and 140 seconds, 
respectively. 
The cross correlation of 
the lightcurves is plotted in the bottom
panel. 
For the 2004 lightcurve determined from aperture photometry
the stellar background estimate is $5.3 \pm 0.2$ mJy which has been subtracted.
For the lightcurve determined
from PSF photometry (separated from S17 and S19 and free from any contribution from
the seeing halo of S2) the specific amount of faint stellar contribution is not
clear, but small, $<$1~mJy. A maximum likelihood value with 2$\sigma$ error bars are 
shown in the bottom panel. 
The 1$\sigma$ error bar is given  in Table 2. 
} 
\end{figure}

\begin{figure} 
\center 
\includegraphics[scale=0.4,angle=0]{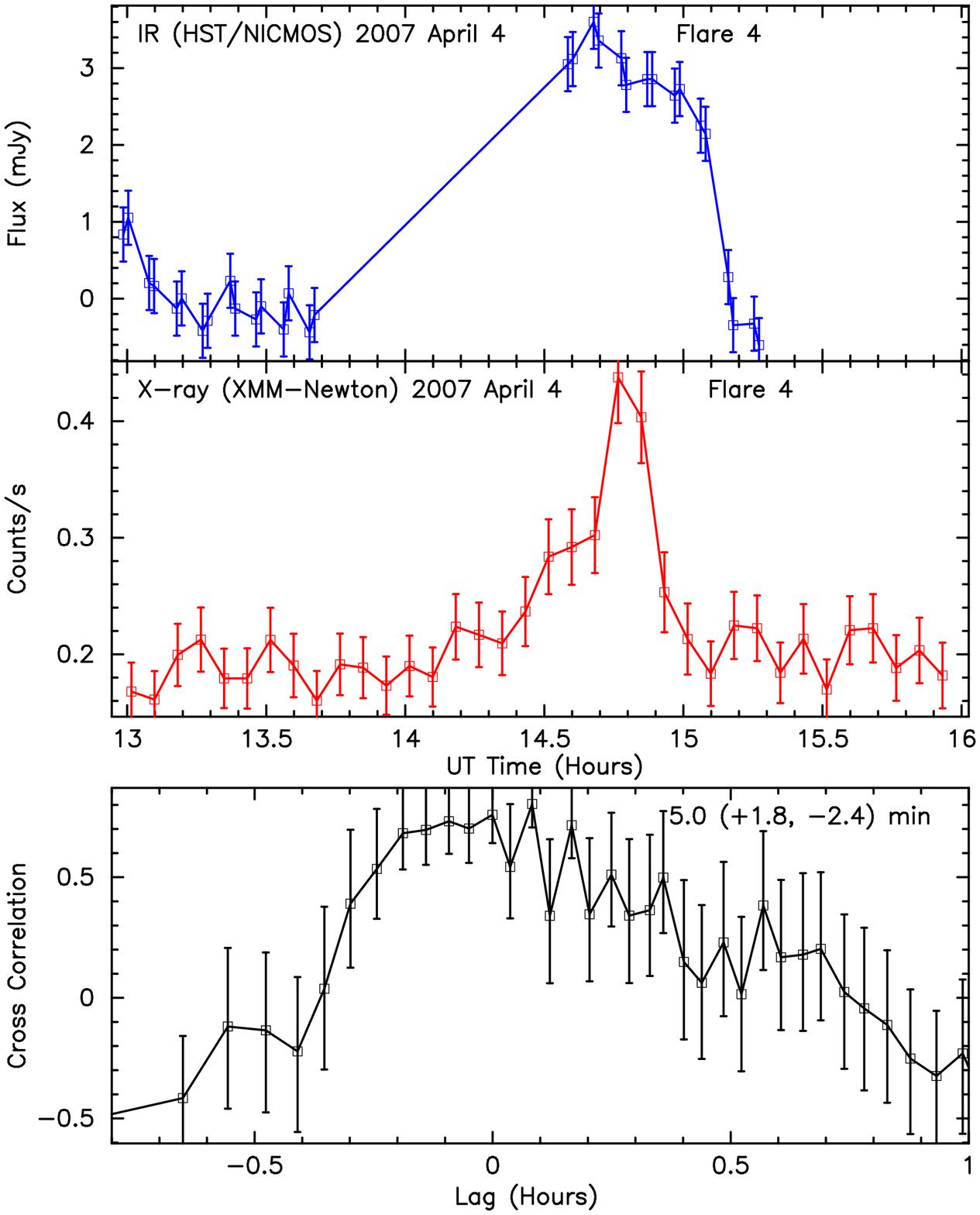} 
\includegraphics[scale=0.4,angle=0]{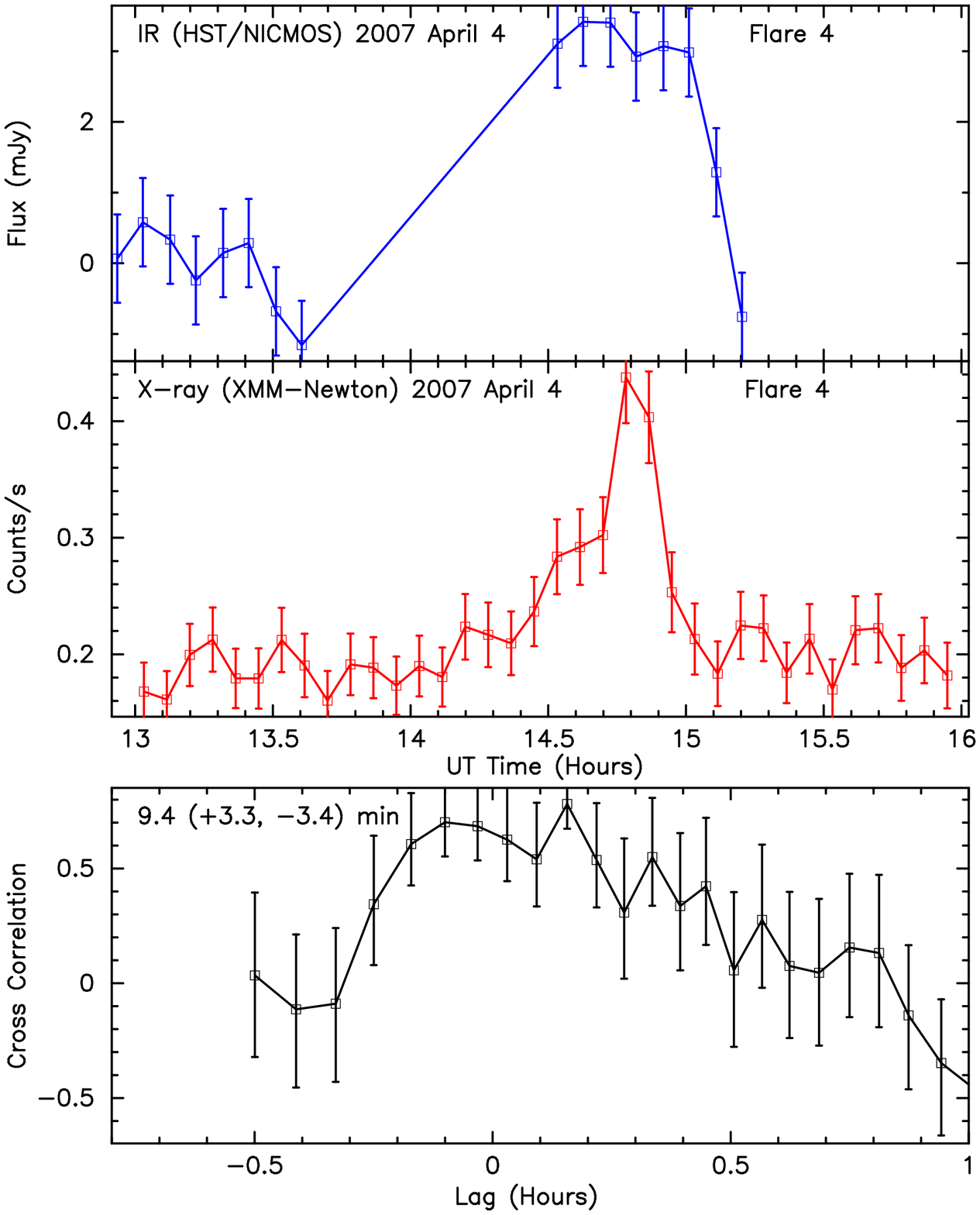}\\
\includegraphics[scale=0.4,angle=0]{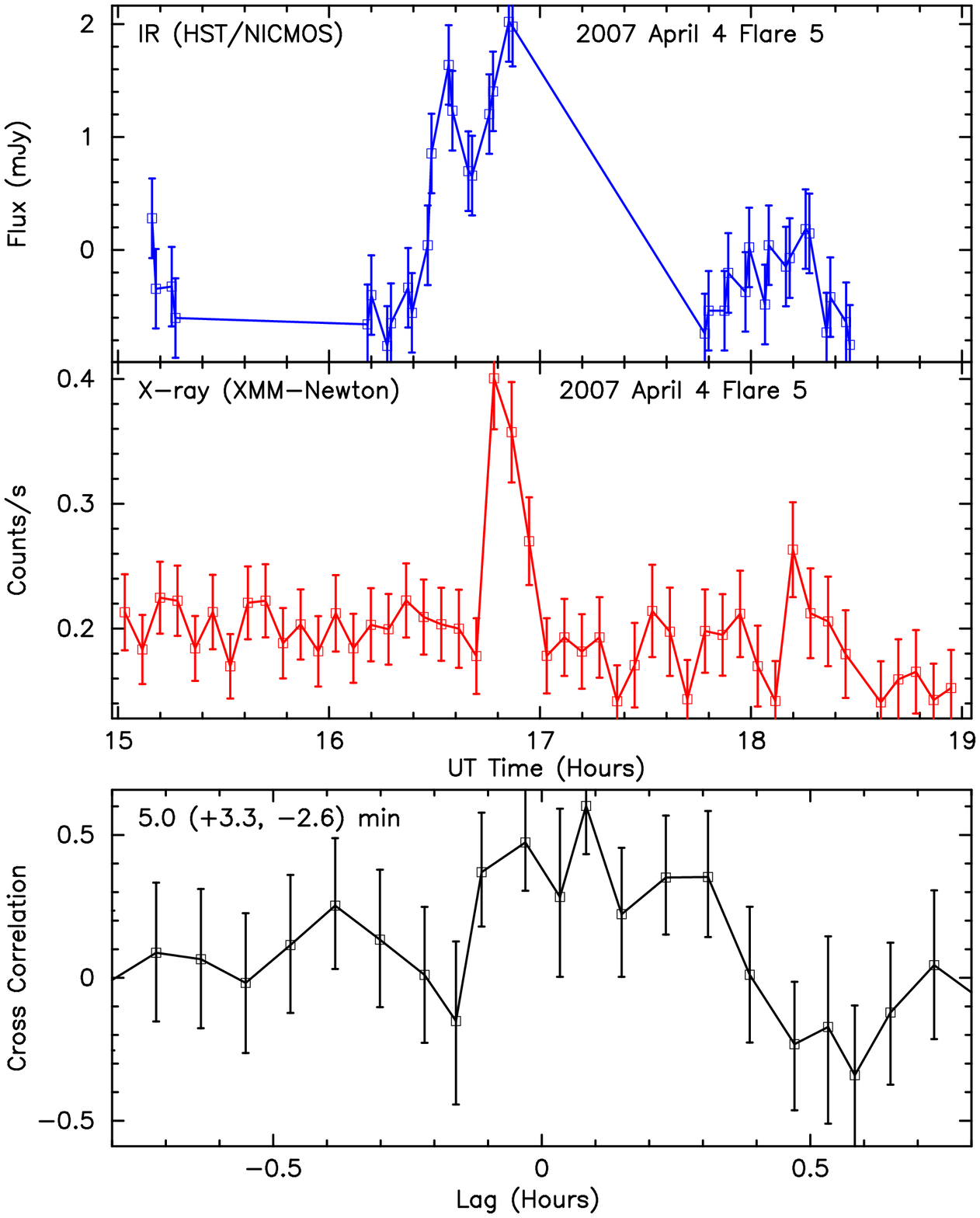} 
\includegraphics[scale=0.4,angle=0]{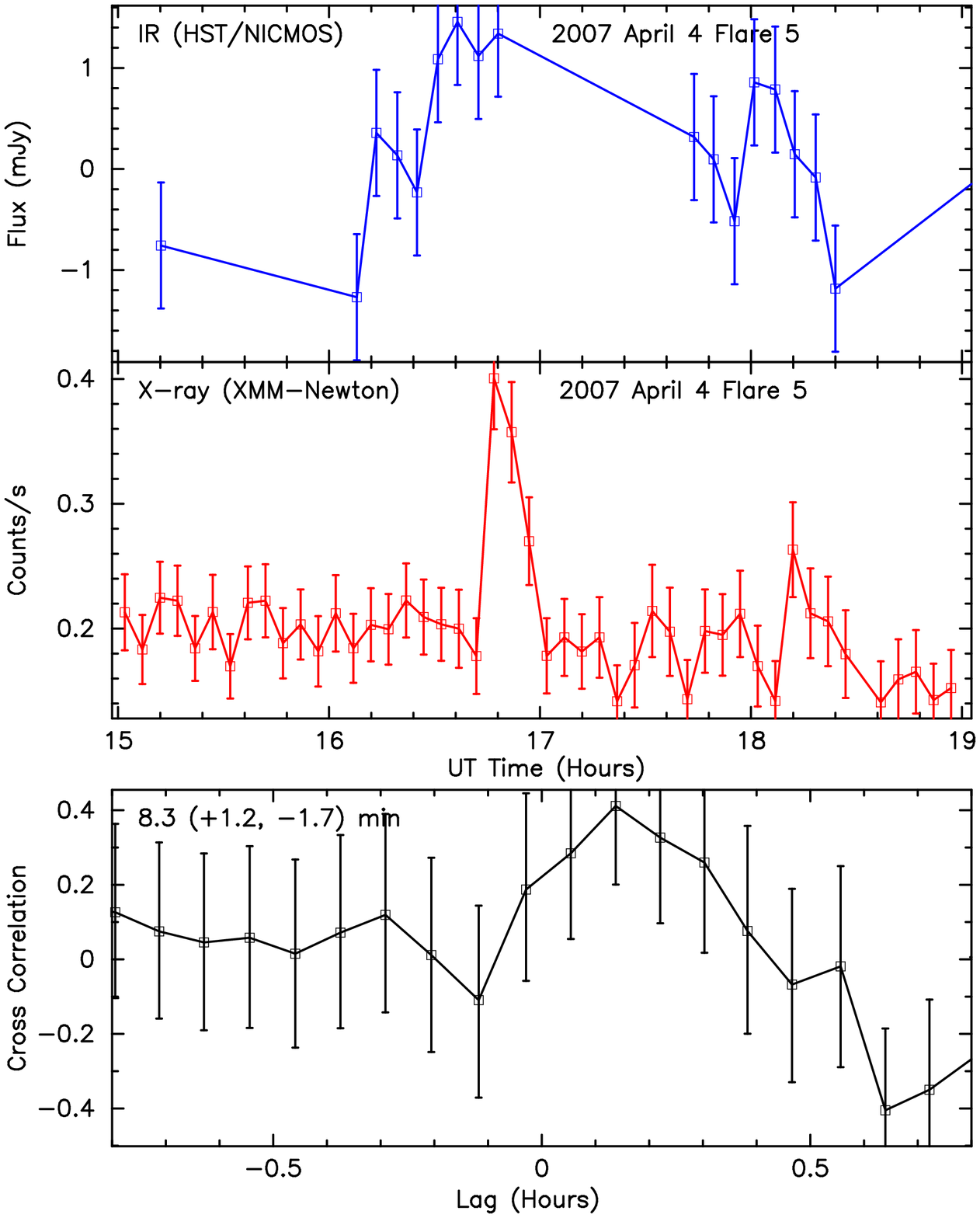} 
\caption{
{\it (a - Top Left)}
The light curves of flare \#4 of 2007 April 4 
in near-IR (1.70$\mu$m) and X-ray (2-10 keV) 
are taken by HST/NICMOS, and XMM/EPIC, with time sampling of 64 and 
300 seconds, respectively. 
{\it (b - Top Right)} The same as (a) except that the 1.45$\mu$m are sampled at 144 second  interval 
to improve the S/N. 
{\it (c - Bottom Left)} The same as (a) except that the light curves of flare \# 5 are displayed at 1.70$\mu$m. 
{\it (d - Bottom Right)} The same as (b) except that the light curves of flare \# 5 are displayed at 1.45$\mu$m. 
The cross correlation and the maximum likelihood values 
with  2$\sigma$ error bars are  shown in  bottom panels.
The 1$\sigma$ error bars for flares \#4 and 5 at 1.70$\mu$m are  given  in Table 2.    
}\end{figure}

\begin{figure}
\center
\includegraphics[scale=0.4,angle=0]{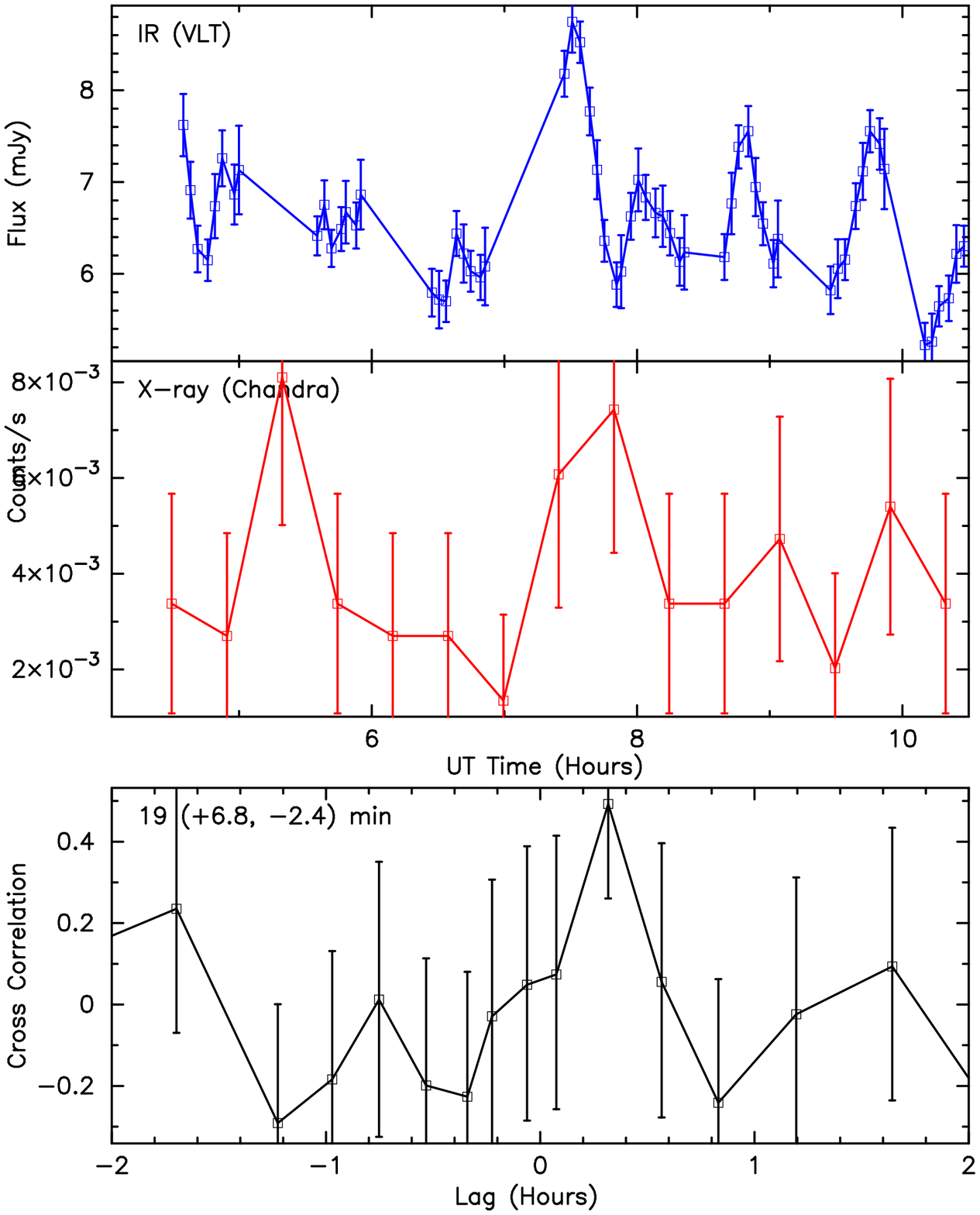}
\includegraphics[scale=0.4,angle=0]{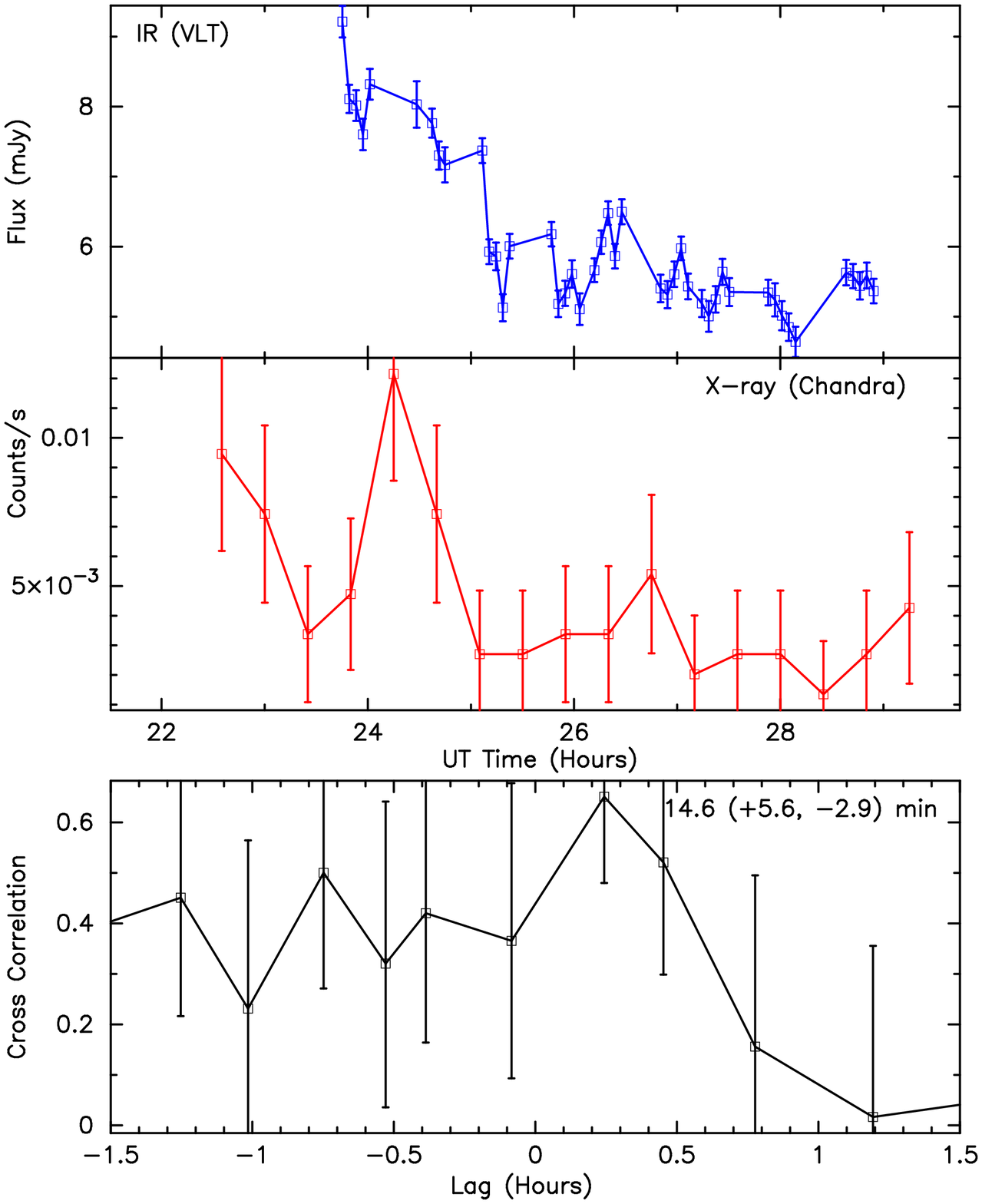}
\includegraphics[scale=0.4,angle=0]{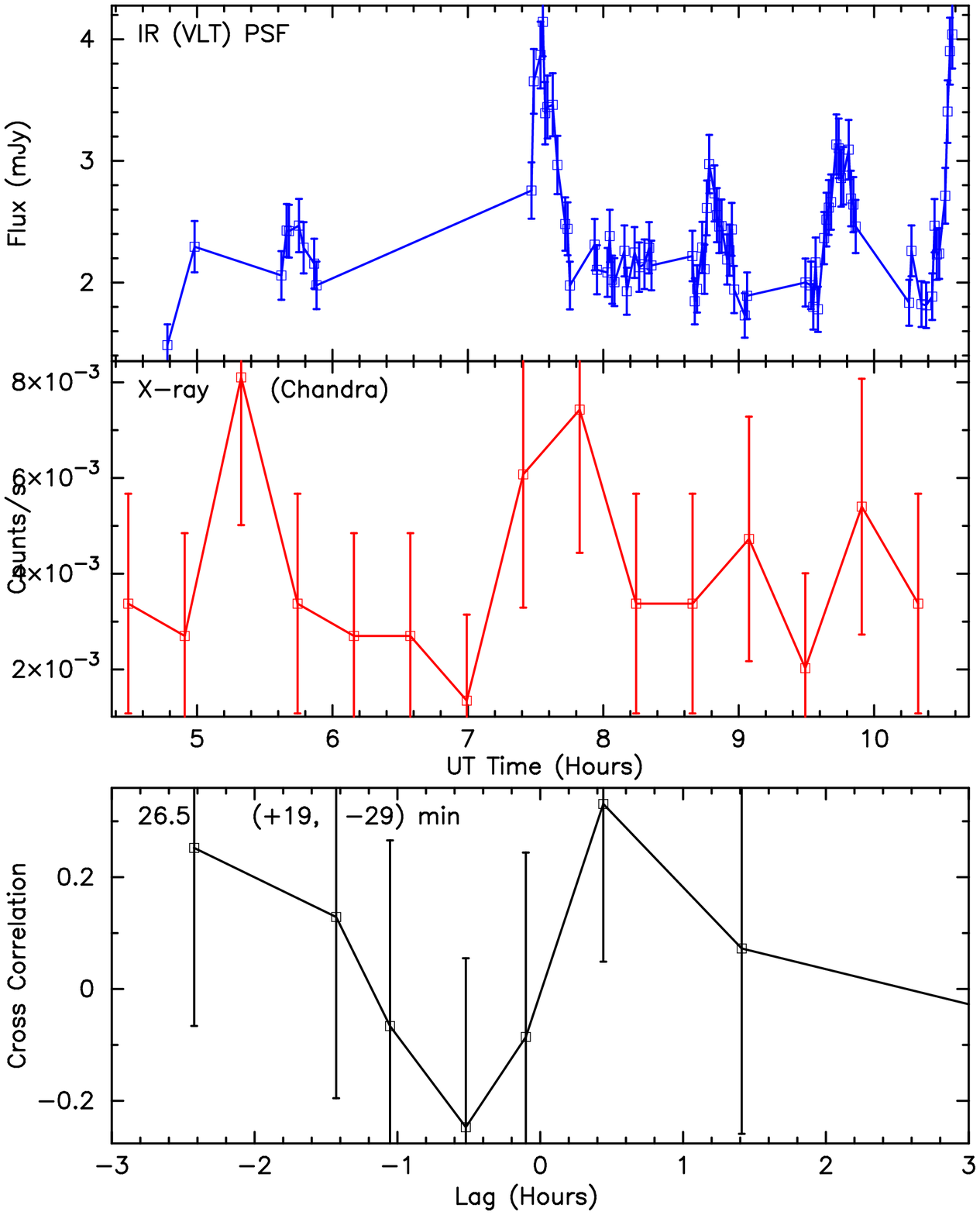}
\includegraphics[scale=0.4,angle=0]{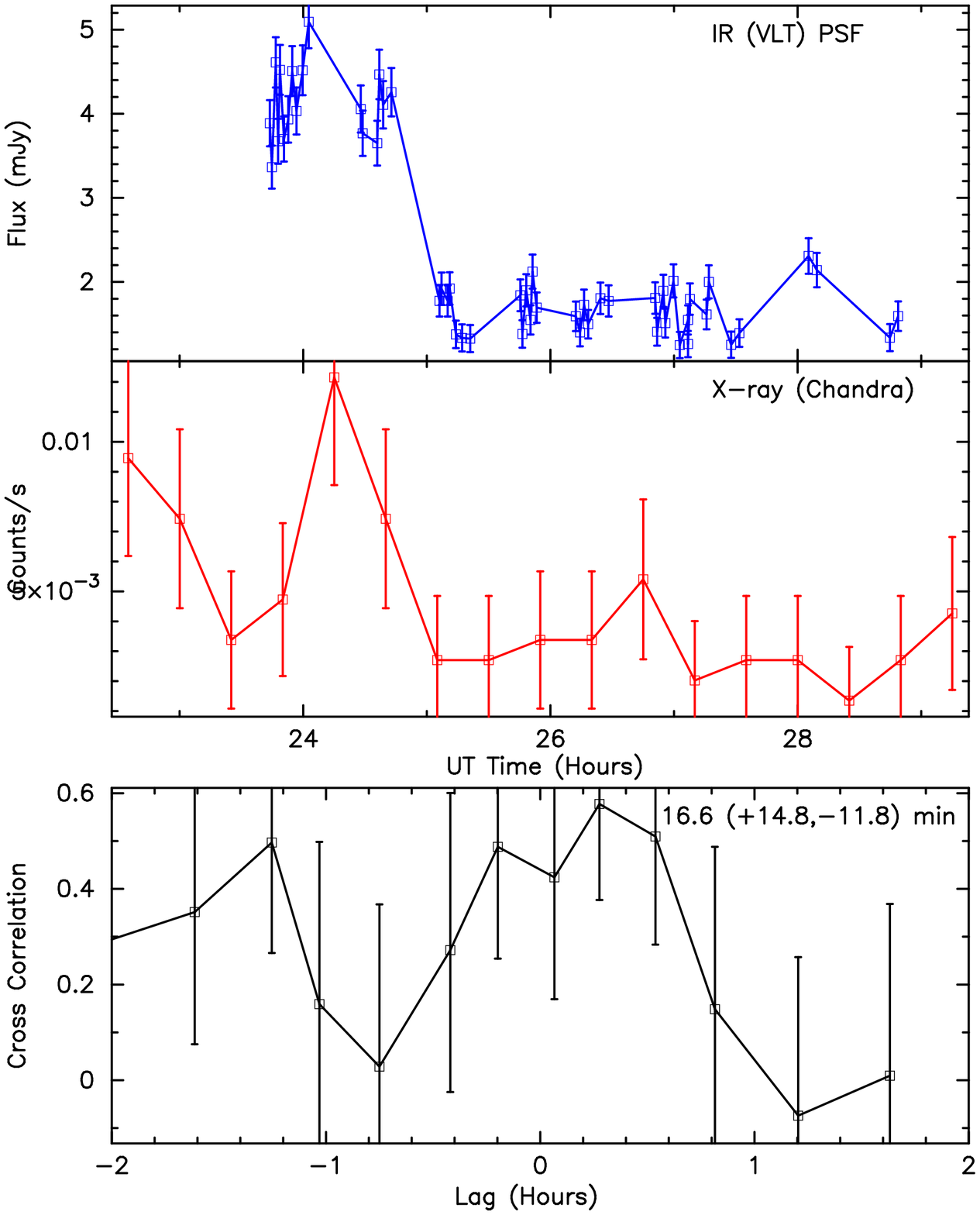}
\caption{ 
{\it (a - Top Left)}
Using aperture photometry technique, 
top two panels show the light curves of Ks-band  (2.2$\mu$m)  
and X-ray (2-8keV) data taken simultaneously with VLT  and 
Chandra  on 2008 May 5.
 The sampling interval for X-ray and near-IR  data are
25  and $\sim2$ minutes, respectively.
The cross correlation of the light curves is plotted in the bottom panel. 
{\it (b - Top Right)}
Similar to (a) except that the data taken on 2008, July 26+27.
{\it (c - Bottom Left)}
Similar to (a) except the light curve of Sgr A* is calibrated using 
PSF photometry technique. 
{\it (d - Bottom Right)}
Similar to (b) except the light curve of Sgr A* is calibrated using 
PSF photometry technique. 
The cross correlation and the maximum likelihood values 
with  2$\sigma$ error bars are  shown in  bottom panels.
The 1$\sigma$ error bars for aperature photometric data are  given  in Table 2.    
} 
\end{figure}

\begin{figure}
\center
\includegraphics[scale=0.4,angle=0]{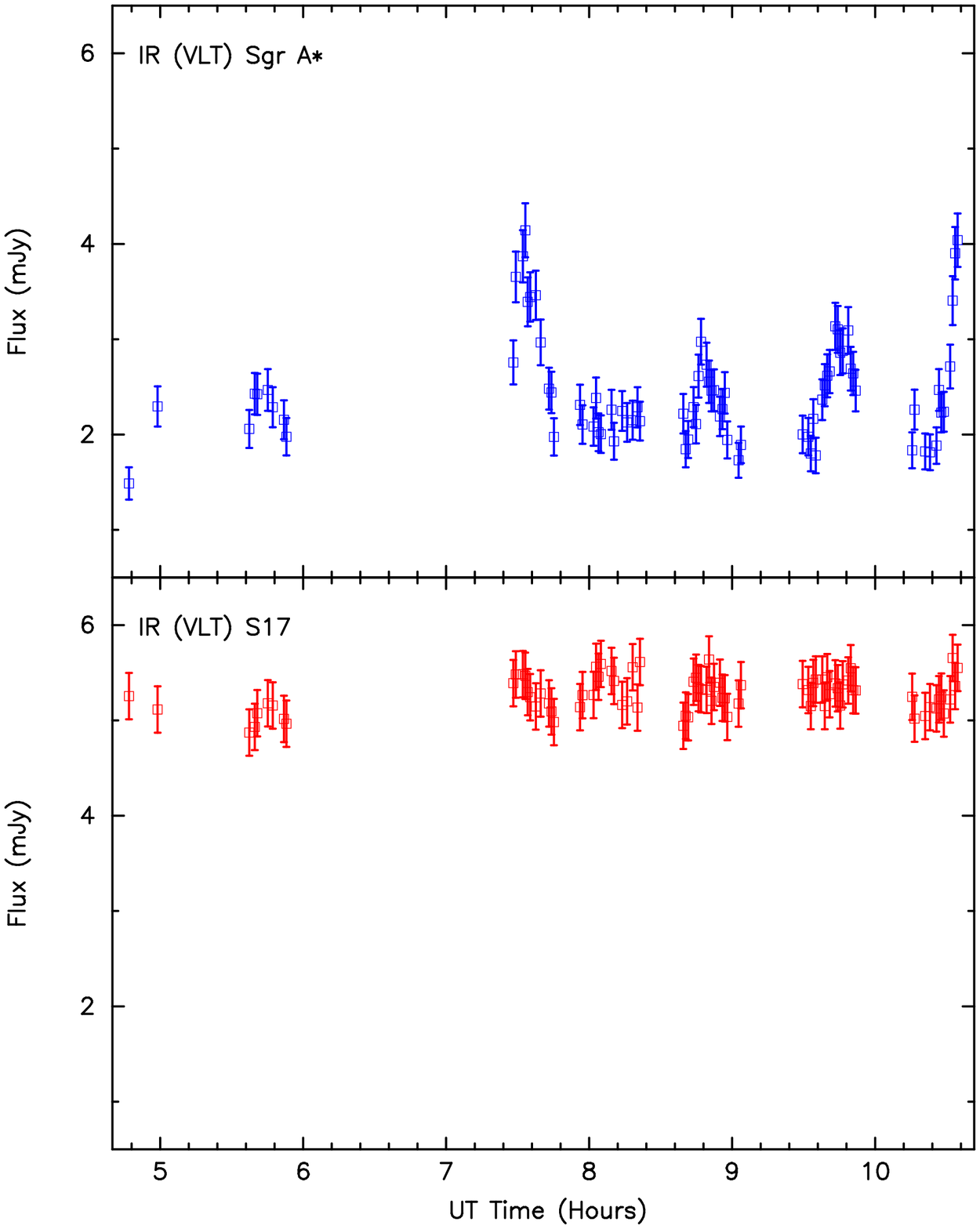}
\includegraphics[scale=0.4,angle=0]{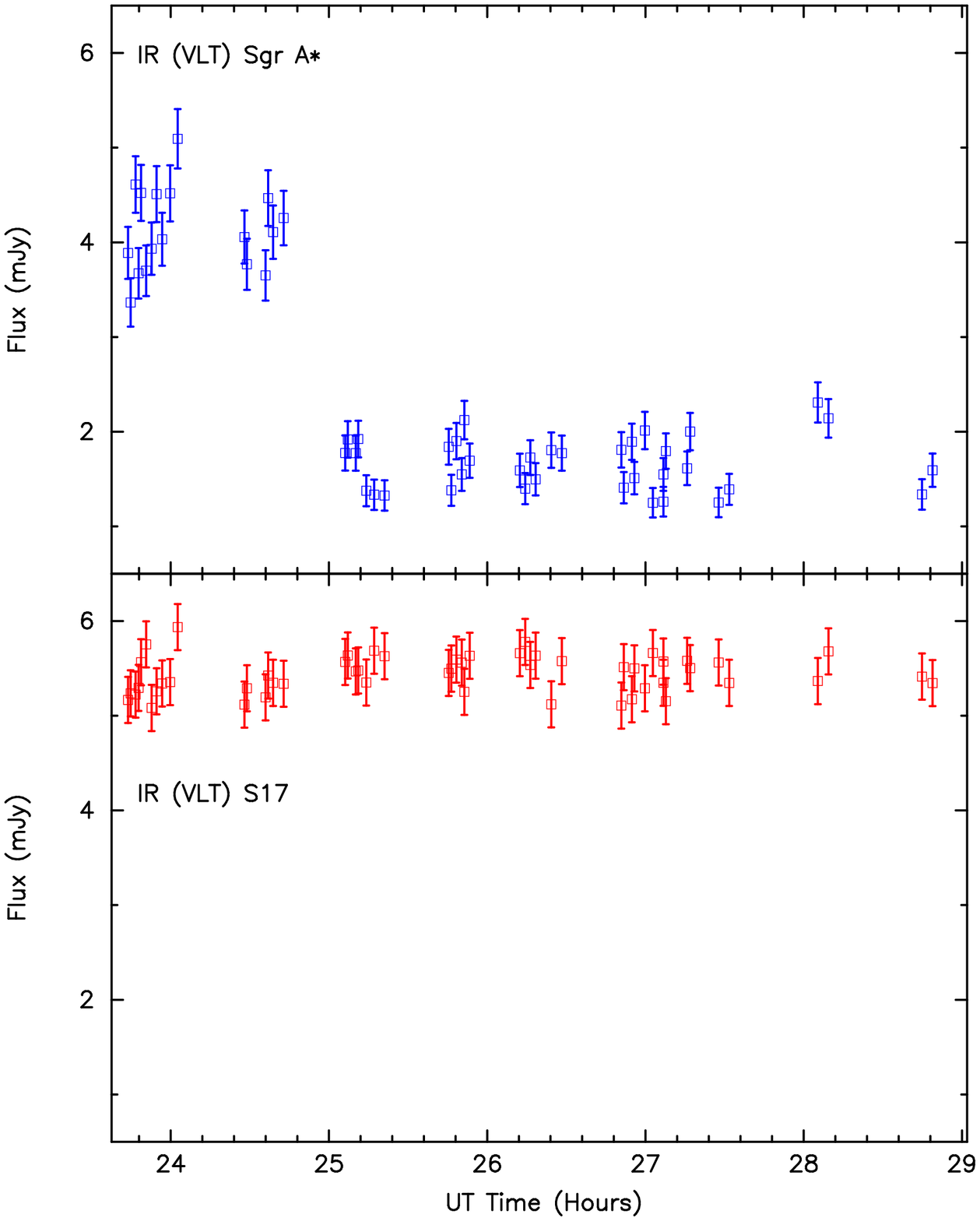}
\caption{ 
{\it (a - Left)}
Top two panels show the light curves of PSF photometrically reduced 
Sgr A* and S17 at 2.2$\mu$m on 2008 May 5. 
{\it (b - Right)}
Similar to (a) except that the data taken on 2008, July 26+27. 
A base level of 3.6 mJy has been subtracted. 
} 
\end{figure}

\begin{figure} 
\center 
\includegraphics[scale=0.5,angle=0]{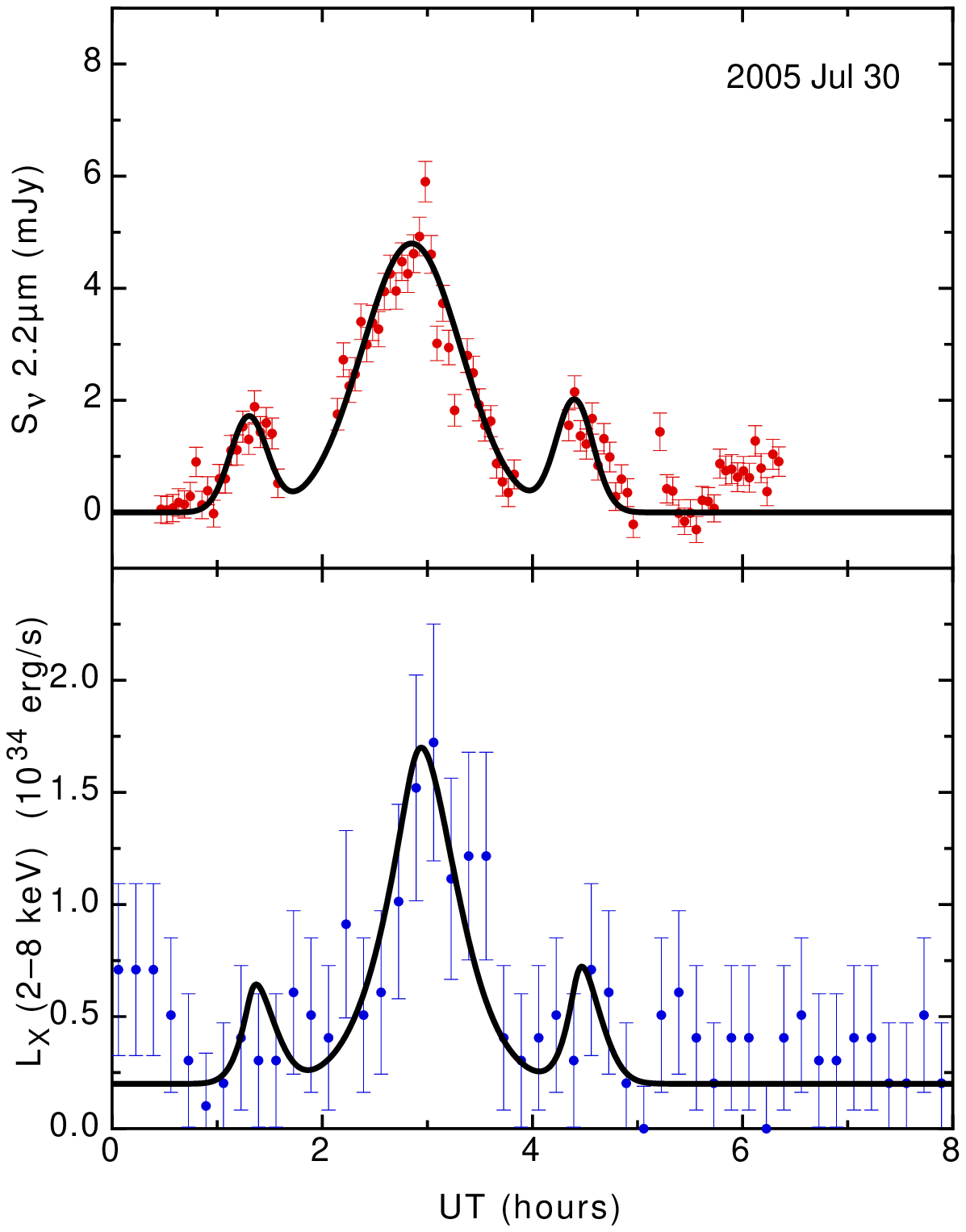} 
\includegraphics[scale=0.5,angle=0]{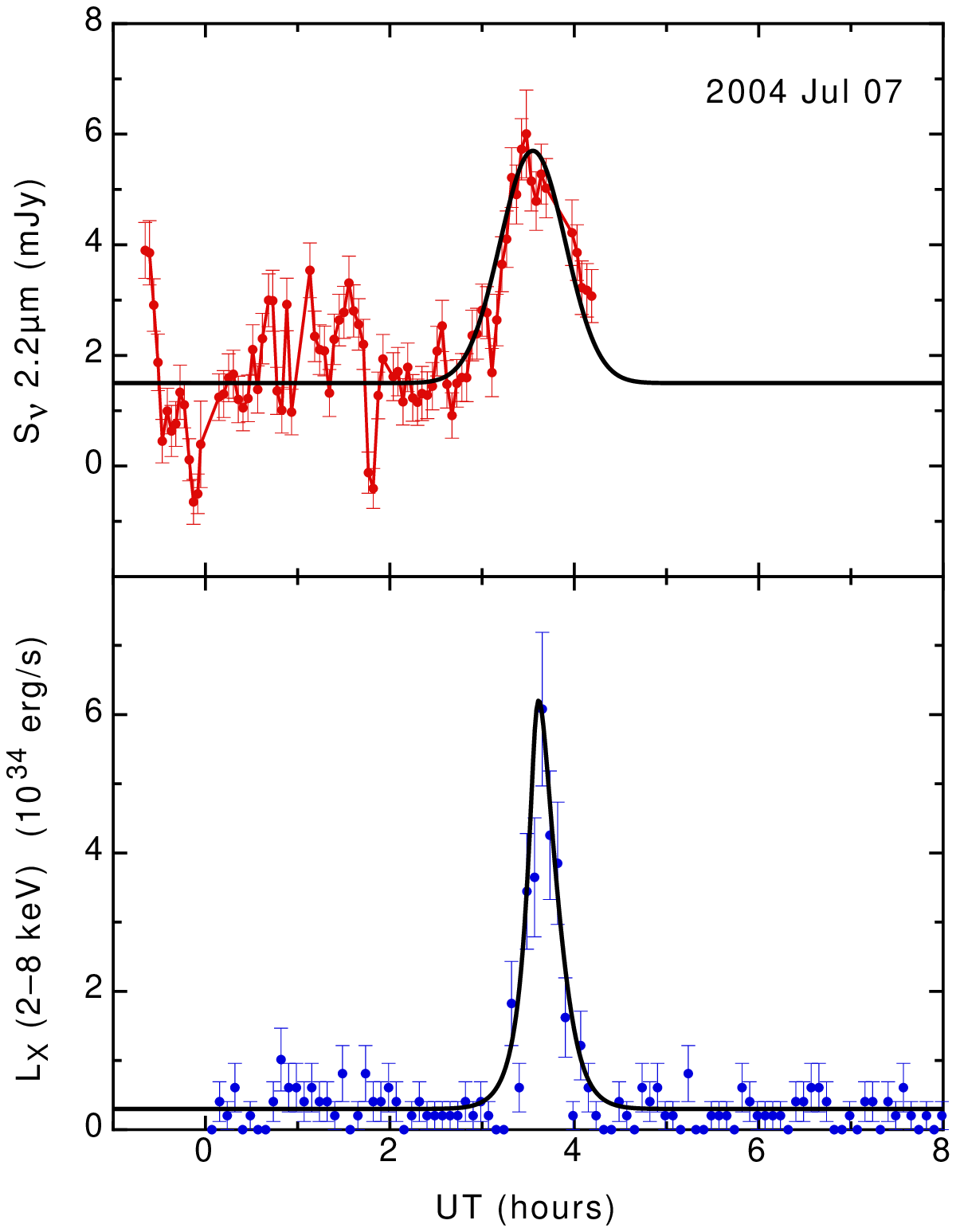}\\ 
\includegraphics[scale=0.5,angle=0]{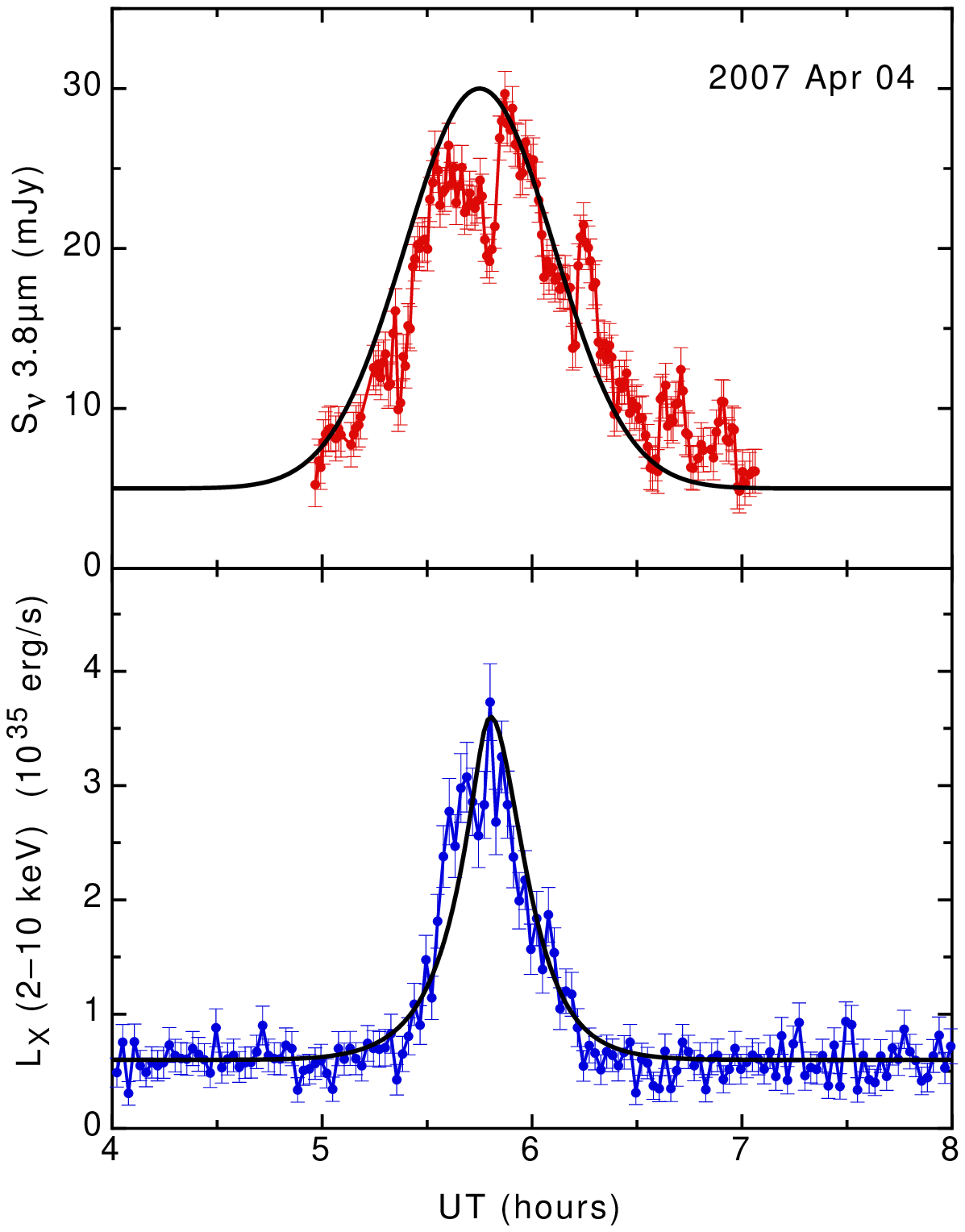} 
\includegraphics[scale=0.5,angle=0]{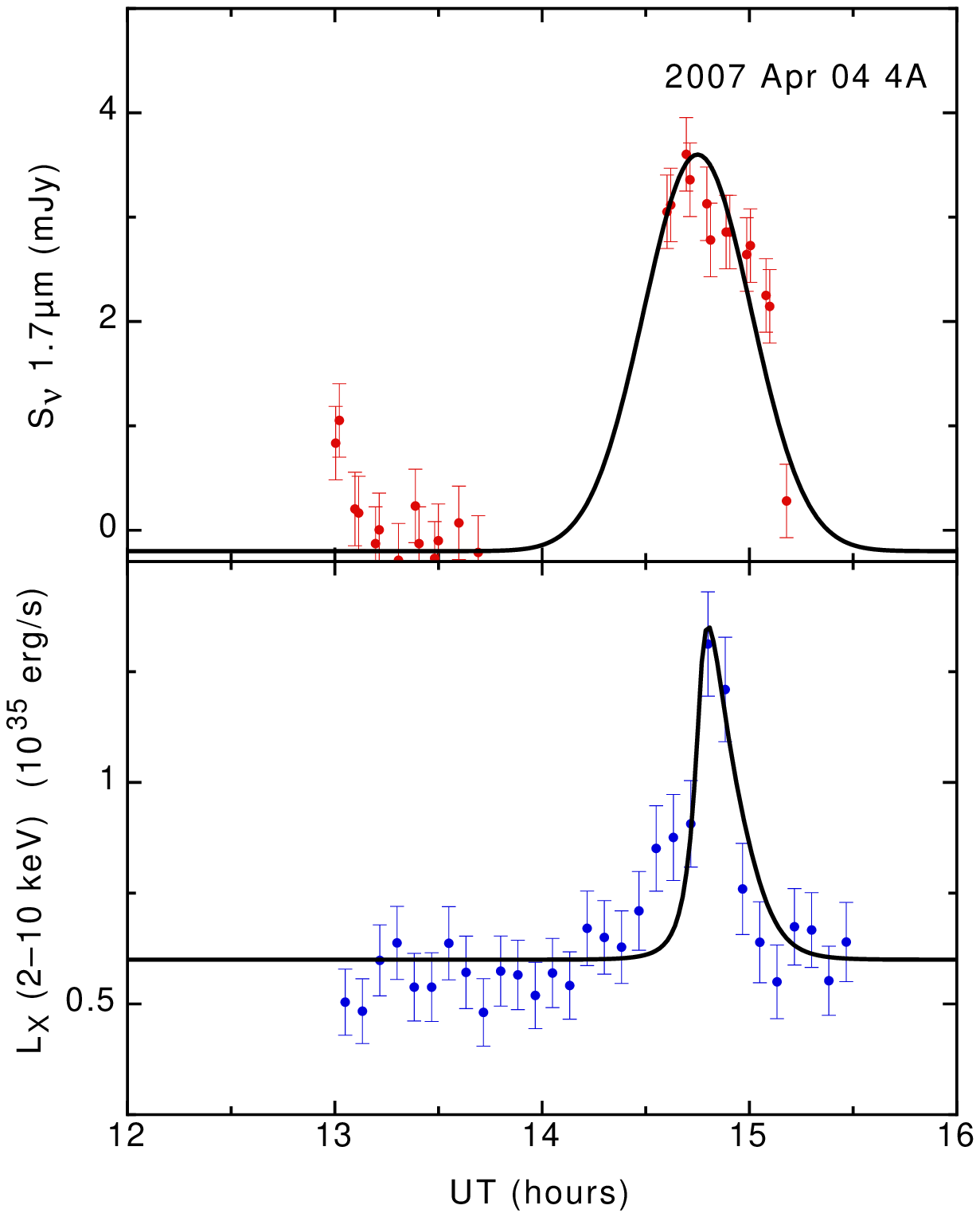} 
\caption{
Adopted near-IR light curves and the
corresponding ICS produced X-ray light curves (solid lines) superimposed on 
the near-IR and X-ray flares observed 
on 2005 July 30 (Top Left),
2004 July 7 (Top Right),
and two flares on 2007, April 4 (Bottom Left and Right), 
 respectively. Near-IR flare data are an input to the ICS model. 
The X-ray and near-IR flare of 
2005 July 30 and 2004 July 7 are taken from Eckart et al. (2006, 2008)
whereas 
the 2007, April 4 data 
are taken from Porquet el al. (2008) and  Dodds-Eden et al. (2009). 
The ICS model parameters are listed  in Table 1. Two weak  flares before and 
after the main flare in the 2005 data in (a) have also been modeled. The 
A possible second    flare
of the 2004 data near 4h UT, as shown in Figure 2, has not been modeled in 
(b). 
} 
\end{figure}

\begin{thebibliography}{99}

\bibitem[ et al.(2000)]{2003ApJ} 


\bibitem[Alexander(1997)]{alexander97} Alexander, T. 1997,  \mnras\,  285, 891

\bibitem[An et al.(2005)]{2005ApJ} 
An, T., Goss, W. M., Zhao, J.-H., Hong, X. Y., Roy, S., Rao, A. P., and 
Shen, Z.-Q. 
2005, \apj,   634,   L49

\bibitem[Baganoff et al.(2003)]{2003ApJ...591..891B} Baganoff, F.~K., et 
al.\ 2003, \apj, 591, 891

\bibitem[Blandford1999]{19993MNRAS} Blandford, R.~D. and Begelman, M.~C.
1999, MNRAS, 303, L1

\bibitem[Dodds-Eden {et~al.}(2009)]{Dodds-Eden} Dodds-Eden, K., Porquet, D., 
Trap, G., Quataert, E., Haubois, X.,  Gillessen, S., Grosso, N., Pantin, 
E., Falcke, H., Rouan, D., Genzel, R., Hasinger, G., Goldwurm, A.,
Yusef-Zadeh, F., Clenet, Y., Trippe, S., Lagage, P.-O., Bartko, 
H., Eisenhauer, F., Ott, T., Paumard, T., Perrin, G., Yuan, F., Fritz, T. 
K. \& Mascetti, L. 2009, ApJ, 698, 676

\bibitem[Dodds-Eden et al.(2010)]{2010ApJ...725..450D} Dodds-Eden, K., 
Sharma, P., Quataert, E., Genzel, R., Gillessen, S., Eisenhauer, F., 
\& Porquet, D.\ 2010, \apj, 725, 450

\bibitem[Dodds-Eden et al. 2011]{dodds-eden2011} 
 Dodds-Eden, K., et al. 2011, ApJ, 728, 37 

\bibitem[Eckart et al. 2006]{eckart2006} {Eckart}, A., {Baganoff}, F.~K., {Sch{\"o}del}, R., {Morris}, M., 
{Genzel}, R.,  {Bower}, G.~C., {Marrone}, D., {Moran}, J.~M., {Viehmann}, T., {Bautz},
  M.~W., {Brandt}, W.~N., {Garmire}, G.~P., {Ott}, T., {Trippe}, S., {Ricker},
  G.~R., {Straubmeier}, C., {Roberts}, D.~A., {Yusef-Zadeh}, F., {Zhao}, J.~H.,
  \& {Rao}, R. 2006{\natexlab{a}}, \aap, 450, 535

\bibitem[Eckart et 
al.(2008)]{Eckart2008} Eckart, A., et al.\ 2008, \aap, 492, 337 

\bibitem[Eckart et 
al.(2009)]{Eckart2009} Eckart, A., et al.\ 2009, \aap, 500, 935


\bibitem[Fritz et al. 2011]{fritz2011} Fritz, T. K.  et al. 2011, in press

\bibitem[Gehrels (1986)]{1986ApJ}
Gehrels, N.  1986,  \apj,   303,   336


\bibitem[Genzel et al.(2010)]{Genzel2010} Genzel, R., Eisenhauer, 
F., \& Gillessen, S.\ 2010, Reviews of Modern Physics, 82, 3121 

\bibitem[Genzel et al.(2003)]{Genzel2003}Genzel, R., Sch\"odel, R., Ott, T., Eckart, A., Alexander, T., Lacombe, F., 
Rouan, D., and Aschenbach, B., 2003,  Nature,   425, 934


\bibitem[{{Ghez} {et~al.}(2005){Ghez}, {Salim}, {Hornstein}, {Tanner}, {Lu},
  {Morris}, {Becklin}, \& {Duch{\^e}ne}}]{ghez05}
{Ghez}, A.~M., {Salim}, S., {Hornstein}, S.~D., {Tanner}, A., {Lu}, J.~R.,
  {Morris}, M., {Becklin}, E.~E., \& {Duch{\^e}ne}, G. 2005, \apj, 620, 744

\bibitem[Gillessen et~al.2009]{gillessen09}
{Gillessen}, S., {Eisenhauer}, F., {Trippe}, S., {Alexander}, T., {Genzel}, R.,
  {Martins}, F., \& {Ott}, T. 2009, \apj, 692, 1075

\bibitem[Goldwurm et al. 2003]{Goldwurm2003ApJ} 
Goldwurm, A., Brion, E., Goldoni, P., Ferrando, P., Daigne, F., 
Decourchelle, A., Warwick, R. S., and Predehl, P. 2003,  \apj,   584,   751

\bibitem[Herrnstein et al.(2004)]{2004AJ}
Herrnstein, R. M., Zhao, J.-H., Bower, G. C., and Goss, W. M. , 2004,  \apj,   127,   3399

\bibitem[Hornstein et~al. 2007]{hornstein07}
  {Hornstein}, S.~D., {Matthews}, K., {Ghez}, A.~M., {Lu}, J.~R., {Morris}, M.,
  {Becklin}, E.~E., {Rafelski}, M., \& {Baganoff}, F.~K. 2007, \apj, 667, 900



\bibitem[Kusunose and  Takahara 2011]{kusunose2011ApJ} Kusunose, M. \&  Takahara, F.  2011, \apj, 726, 54 

\bibitem[L. et al. 2003]{lenzen2003}Lenzen, R.  et al. 22003, Proc. SPIE, 4841, 944 

\bibitem[Liu et al.(2002)]{2002ApJ} 
Liu, S. and Melia, F. 2002, \apj,   566,   L77

\bibitem[Liu et al.(2004)]{liu04} {Liu}, S., {Petrosian}, V., \& {Melia}, F. 2004, \apjl, 611, L101

\bibitem[Loeb 
\& Waxman(2007)]{2007JCAP...03..011L} 
Loeb, A., \& Waxman, E.\ 2007, \jcap, 3, 11 

\bibitem[Markoff et al.(2001)]{2001AA} 
Markoff, S., Falcke, H., Yuan, F., and Biermann, P. L. 2001,  A\&A, 379,   L13

\bibitem[marrone et al. 2003]{marrone2006}
Marrone, D. P., Moran, J. M., Zhao, J.-H., and Rao, R. 2006,  Journal of Physics Conference Series,   54,   354

\bibitem[]{} Marrone, D. P., Baganoff, F. K., Morris, M. R., Moran, J. M., Ghez, A. M., Hornstein, S. D., et al.
2008, ApJ, 682, 373

\bibitem[Melia \& Falcke(2001)]{melia01} Melia, F. \& Falcke, H. 2001, \araa, 39, 309

\bibitem[Miyazaki et al.(2004)]{2004ApJ} 
Miyazaki, A., Tsutsumi, T., and Tsuboi, M., 2004,  \apj,   611,   L97

\bibitem[Muno et al.(2005)]{Muno2005}
Muno, M. et al.  2005, \apj, 633, 228 


\bibitem[M. et al. 2009]{M2009}
Mos\'cibrodzka, M., Gammie, C. F., Dolence, J. C., Shiokawa, H. \& Leung, P. K. 2009,
ApJ, 706, 497

\bibitem[Porquet et~al. 2008]{porquet08}
{Porquet}, D., {Grosso}, N., {Predehl}, P., {Hasinger}, G., {Yusef-Zadeh}, F.,
  {Aschenbach}, B., {Trap}, G., {Melia}, F., {Warwick}, R.~S., {Goldwurm}, A.,
  {B{\'e}langer}, G., {Tanaka}, Y., {Genzel}, R., {Dodds-Eden}, K., {Sakano},  
  M., \& {Ferrando}, P. 2008, A\&A, 488, 549

\bibitem[Reid (1993)]{reid93} Reid, M. J. 1993, ARAA, 31,  345

\bibitem[Reid \& Brunthaler (2004)]{reid04}
{Reid}, M.~J., \& {Brunthaler}, A. 2004, \apj, 616, 872


\bibitem[Rousset et al. 2003]{Rousset2003} 
Rousset, G., et al. 2003, Proc. SPIE, 4839, 140

\bibitem[Rybicki and Lightman]{Rybicki1986} Rybicki, G. B. \& Lightman, Alan P. 1986, 
Chapter 7.5,  Radiative Processes in Astrophysics, 
ISBN 0-471-82759-2. Wiley-VCH 

\bibitem[]{} Sabha, N., Witzel, G., Eckart, A., Buchholz, R. M., Bremer, M., Gie\ss\"ubel, R., et al. 2010, \aap,
512, A2

\bibitem[]{} Sazonov, S., Sunyaev, R., and Revnivtsev, M. 2011,  ArXiv e-prints (1108.2778)



\bibitem[]{} Trap, G., Goldwurm, A., Dodds-Eden, K., Weiss, A., Terrier, R., Ponti, G., et al. 2011, \aap, 528,
A140


\bibitem[Wardle(2011)]{2011ASPC..439..450W} Wardle, M.\ 2011, Astronomical 
Society of the Pacific Conference Series, 439, 450 


\bibitem[Yuan et al.(2003)]{2003ApJ...598..301Y} Yuan, F., Quataert, E., 
\& Narayan, R.\ 2003, \apj, 598, 301 

\bibitem[Yuan et al.(2004)]{2004ApJ} 
Yuan, F., Quataert, E., and Narayan, R.  2004, \apj,  606,   894

\bibitem[Yusef-Zadeh et al.(2006a)]{2006ApJ...644..198Y} Yusef-Zadeh, F., et 
al.\ 2006a, \apj, 644, 198 

\bibitem[Yusef-Zadeh et al. 2009]{fyz2009}Yusef-Zadeh, F., Bushouse, H., 
Wardle, M., Heinke, C., Roberts, D. A., et al. 2009, ApJ, 706, 706


\bibitem[Yusef-Zadeh et al.(2006b)]{2006ApJ} 
Yusef-Zadeh, F., Roberts, D., Wardle, M., Heinke, C. O., and Bower, G. C., 2006b,  \apj,   650,   189

\bibitem[Zhao et al.(2001)]{2001ApJ}
Zhao, J.-H., Bower, G. C., and Goss, W. M.  2001, ApJ, 547, L29

\end{thebibliography}
\end{document}